\documentclass[10pt]{iopart}
\usepackage{amsbsy,amsfonts,amssymb,graphics,epsfig,float,mathrsfs,amsthm,braket}
\usepackage{rotating,booktabs,array,tabularx}

\usepackage{ifsym}
\usepackage{psfrag}
\usepackage[normalem]{ulem}
\usepackage{color}
\usepackage[11pt]{moresize}

\newcommand \beq{\begin{equation}}
\newcommand \eeq{\end{equation}}
\newcommand \beqn{\begin{equation*}}
\newcommand \eeqn{\end{equation*}}

\begin{document}

\title{Delayed pull-in transitions in overdamped MEMS devices}
\footnote{This is an author-created, un-copyedited version of an article accepted for publication/published in the Journal of Micromechanics and Microengineering. IOP Publishing Ltd is not responsible for any errors or omissions in this version of the manuscript or any version derived from it. The Version of Record is available online at https://doi.org/10.1088/1361-6439/aa9a70.}
\author{Michael Gomez, Derek E.~Moulton, Dominic Vella}
\address{Mathematical Institute, University of Oxford, Woodstock Rd, Oxford, OX2 6GG, UK}
\ead{dominic.vella@maths.ox.ac.uk}

\begin{abstract}
We consider the dynamics of overdamped MEMS devices undergoing the pull-in instability. Numerous previous experiments and numerical simulations have shown a significant increase in the pull-in time under DC voltages close to the pull-in voltage. Here the transient dynamics slow down as the device passes through a meta-stable or bottleneck phase, but this slowing down is not well understood quantitatively. Using a lumped parallel-plate model, we perform a detailed analysis of the pull-in dynamics in this regime. We show that the bottleneck phenomenon is a type of critical slowing down arising from the pull-in transition. This allows us to show that the pull-in time obeys an inverse square-root scaling law as the transition is approached; moreover we determine an analytical expression for this pull-in time. We then compare our prediction to a wide range of pull-in time data reported in the literature, showing that the observed slowing down is well captured by our scaling law, which appears to be generic for overdamped pull-in under DC loads. This realization provides a useful design rule with which to tune dynamic response in applications, including state-of-the-art accelerometers and pressure sensors that use pull-in time as a sensing mechanism. We also propose a method to estimate the pull-in voltage based only on data of the pull-in times. 

\end{abstract}

\maketitle
\ioptwocol

\section{Introduction}
\label{sec:intro}
Electrostatic actuation is the most common actuation mechanism in microelectromechanical systems (MEMS): it offers rapid response times, low power consumption and compatibility with existing circuit technology \cite{batra2007,krylov2008}. However, the operation of electrostatic devices is limited by the `pull-in' instability \cite{pelesko2002} in which an elastic structure suddenly collapses towards a nearby electrode when a critical voltage is exceeded. Pull-in can result in failure via short circuit or stiction between components. For this reason, studies have traditionally focussed on the stability of devices under a combination of electrostatic and mechanical restoring forces, with a view to developing methods that extend the operating range of a device prior to pull-in \cite{zhang2014}.

More recently, pull-in has been identified as a useful instability for smart applications. For example, the critical voltage required to pull-in is commonly used in mass sensing applications \cite{younis2009} and to estimate material parameters such as the elastic modulus \cite{osterberg1997}. The dynamics of the pull-in transition is also becoming the basis of many MEMS devices. In these scenarios pull-in is allowed to proceed safely (e.g.~by limiting the displacement of the structure to prevent contact between components), enabling fast motions and large relative displacements to be generated in a reproducible way. For example, microvalves make use of the collapsed state to block off fluid flow in microchannels \cite{desai2012}, and microswitches harness pull-in to rapidly switch between two remote configurations, corresponding to distinct `off' and `on' states \cite{nguyen1998,larose2010,deng2017}. In these applications, an understanding of the pull-in dynamics is essential since it determines the switching time of the device. 

The time taken to pull-in can itself be used as a sensing mechanism: the relationship between the pull-in time of a microbeam and the ambient air pressure has been proposed as a pressure sensor \cite{gupta1997}, while high-resolution accelerometers make use of the sensitivity of parallel-plate actuators to external acceleration \cite{rocha2004a,dias2011,dias2015}. In these applications, unlike microswitches and other actuators,  it is not desirable to simply minimize the pull-in time. Instead, the device is operated at voltages very close to the pull-in transition, where the transient dynamics are observed to slow down considerably. Crucially, this slowing down is highly sensitive to ambient conditions, including external forces, and so has widespread potential to enable high-resolution, low-noise measurements to be made. Using pull-in time as a sensing mechanism also offers the advantage that the device may be integrated in standard circuit technology, so that commercially available micromachining processes can be used \cite{dias2011}.

The slowing down observed in parallel-plate actuators has been attributed to a  `bottleneck' or `meta-stable' phase that dominates the dynamics during pull-in, characterized by a temporary balance between electrostatic and mechanical restoring forces \cite{rocha2004a}: as the net force on the structure is very small, it evolves slowly and the pull-in time is large. However, a quantitative understanding of this bottleneck phenomenon is still lacking, despite the obvious importance of this regime in the operation of many MEMS devices. In particular, it is not clear how the duration of the bottleneck (and hence the pull-in time) scales with the applied voltage, the external acceleration, and the material parameters of the system. 

\subsection{Models of pull-in dynamics}
It is well known that pull-in is initiated by a saddle-node (fold) bifurcation: the equilibrium state away from collapse ceases to exist  above a critical voltage (without first becoming unstable), so the system must pull-in to remain in equilibrium. This means that a standard linear stability analysis cannot be used to study the transient dynamics --- there is no unstable equilibrium base state from which the system evolves. Most studies therefore adopt a purely numerical or experimental approach. 

For devices operating in atmospheric conditions, fluid damping (arising in the squeeze film when the air gap between components becomes very small) has been identified as playing a dominant role \cite{missoffe2008}. A large number of studies have therefore focussed on generating macromodels, i.e.~reduced-order models that couple deformations of the structure to realistic models of the squeeze film damping, including compressibility and rarefaction effects. These macromodels are then used to reduce the computational cost of simulating MEMS devices during pull-in (see \cite{batra2007} and \cite{nayfeh2005} and references therein). Bottleneck phenomena have also been described in a number of macromodel simulations of microbeam actuators \cite{gretillat1997,hung1999,younis2003,nayfeh2007}; slowing down appears to be a generic feature of the dynamics when the system is operated near the pull-in transition, though this has not been explored further. More recent studies instead address the dynamic stability of MEMS resonators under a combination of AC and DC loads \cite{nayfeh2007,zaitsev2012}, the effects of geometric nonlinearities due to large displacements \cite{chaterjee2009}, contact bouncing \cite{larose2010,mccarthy2002}, and modelling structures that possess natural curvature where snap-through buckling can occur alongside pull-in \cite{krylov2008,das2009,krylov2010}; for a review see \cite{zhang2014}. 

Few analytical results concerning pull-in dynamics are available. While general bounds on the pull-in time have been obtained  \cite{flores2003}, these bounds are not very tight  and do not give insight into possible slowing down behaviour close to pull-in. In the case of underdamped, inertia-driven systems, some progress has been made.  For devices operating at very low ambient pressures, inertial effects can cause the critical voltage at pull-in (the dynamic pull-in voltage) to decrease compared to that obtained when the voltage is quasi-statically varied (the static pull-in voltage) \cite{nielson2006}. Using energy methods, scaling laws for the pull-in time have been derived for parallel-plate actuators \cite{leus2008} and extracted for more complex devices such as microbeams using lumped-parameter models \cite{joglekar2011}. The key result is that the pull-in time, $t_{\mathrm{PI}}$, scales logarithmically with the difference between the applied voltage and the pull-in voltage, $\Delta V > 0$: we have that 
\beq
t_{\mathrm{PI}} \propto \log(1/\Delta V),
\label{eqn:UnderDamped}
\eeq as $\Delta V \to 0$. As $\Delta V$ decreases, the pull-in time therefore increases rapidly, until eventually mechanical noise limits the response.

Due to its simplicity, the  scaling law (\ref{eqn:UnderDamped}) offers a useful design rule to tune dynamic response in applications: only a small number of runs are needed to extract the appropriate pre-factor in the scaling law to make further predictions. The need to perform parameter sweeps that at each stage involve detailed simulations can then be eliminated. However, no corresponding scaling law has been found for overdamped systems, despite the fact that (i) many MEMS devices operate in this regime \cite{rocha2004a} and (ii) there is a clear need for such a design rule as devices continue to scale down and grow in complexity \cite{loh2012}.  Obtaining such a scaling law analytically is the primary objective of this paper. 

\subsection{A scaling law for overdamped pull-in}
To explore the possibility of a scaling law analogous to (\ref{eqn:UnderDamped}) but applicable to overdamped devices, we have assembled a large range of data for pull-in times reported in the literature. We focus on results for devices operating at (or near) atmospheric pressure only; we do not consider data for pull-in times in vacuum where inertial effects are important. We consider parallel-plate and microbeam devices, incorporating results from both experiments and  dynamic simulations. This includes data where the actuation voltage is varying while the external acceleration is zero, as well as data from pull-in time accelerometers where the actuation voltage is fixed but the external acceleration is varied. A summary of the conditions for each data set are provided later in tables \ref{table:parallelplate}--\ref{table:microbeam}. In all cases examined, the pull-in times are measured from the point of application of a step DC voltage (stepped from zero). Where data is only available graphically, we have extracted the values using the WebPlotDigitizer (arohatgi.info/WebPlotDigitizer).

\begin{figure*}
\centering
\includegraphics[width = 1.4\columnwidth]{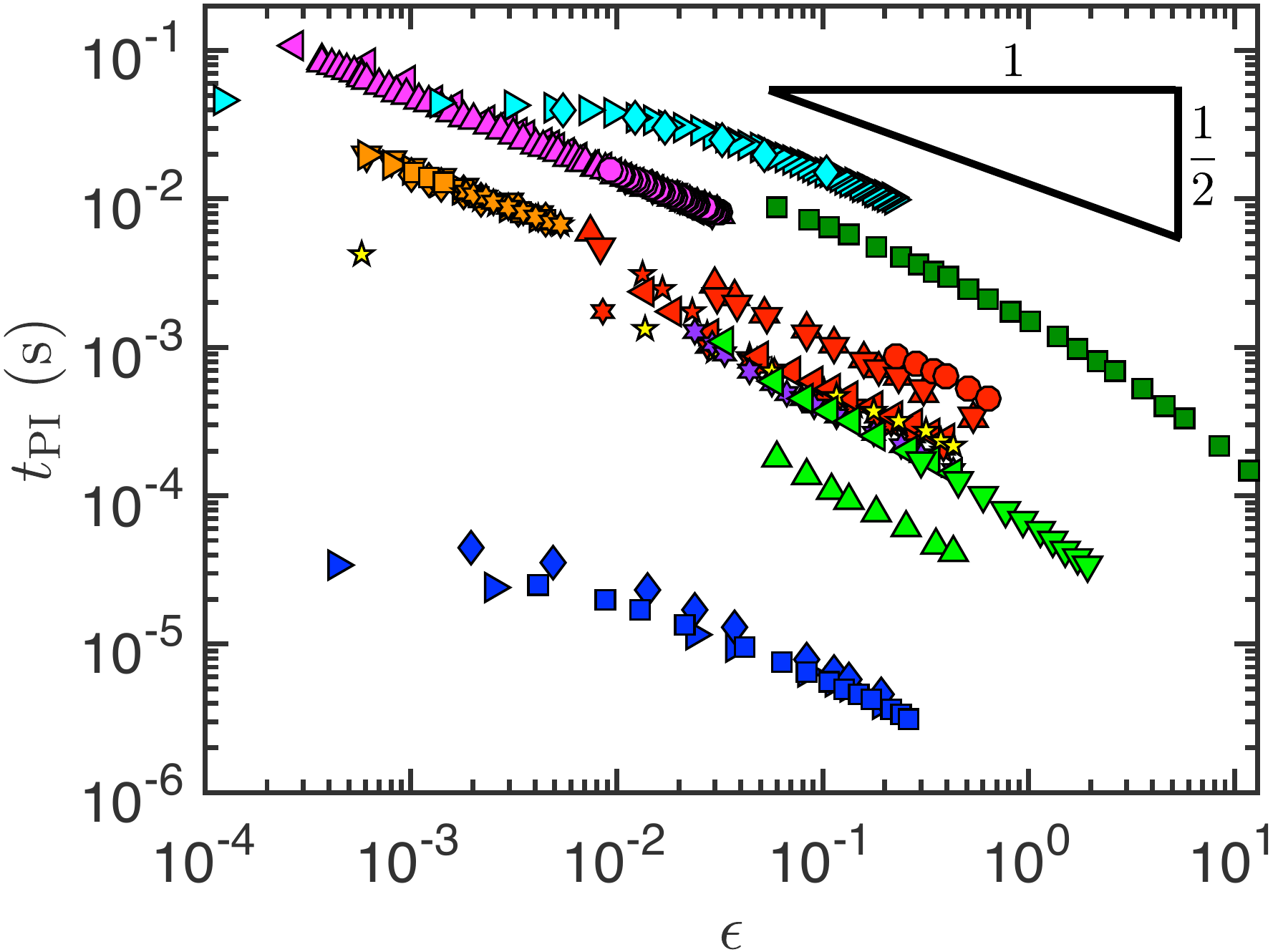} 
\caption{Pull-in times of parallel-plate and microbeam devices under step DC loads reported in the literature. In total, $27$ sets of data from $9$ different references are included, indicated by different symbol shape and colour (details and legend are provided in tables \ref{table:parallelplate}--\ref{table:microbeam}).}
\label{fig:slowingdown}
\end{figure*}

For each measurement, we use the reported values of the pull-in voltage to compute the normalized distance to the pull-in transition, which we denote by $\epsilon$. In particular, in the case of zero external acceleration $\epsilon = (V/V_{\mathrm{SPI}})^2-1$ with $V$  the applied voltage and $V_{\mathrm{SPI}}$ the static pull-in voltage. The results are shown on logarithmic axes in figure \ref{fig:slowingdown}, where different symbols are used to indicate different data sets (i.e.~where the properties of the actuator are varied), and data from different references are distinguished using different colours. We observe that the pull-in time increases as $\epsilon$ decreases in a  systematic way. Very close to the pull-in transition, the dynamics become highly sensitive to the precise value of $\epsilon$: the pull-in time may increase by over an order of magnitude within a very narrow range of $\epsilon$. This is the bottleneck regime in which the dynamics of pull-in are dramatically slowed down.

This delay behaviour is reminiscent of the critical slowing down observed near saddle-node bifurcations in a range of physical systems, such as elastic snap-through \cite{Gomez2017}, phase transitions \cite{chaikin}, and the switching of charge density waves \cite{strogatz1989}. In these systems, the remnant or `ghost' of the saddle-node bifurcation continues to attract trajectories that are nearby in parameter space, producing a bottleneck whose duration generically increases with decreasing distance from the bifurcation  \cite{strogatz2015}. The detailed scaling of the duration of this bottleneck phase depends on the importance of inertia: a scaling $\propto\epsilon^{-1/2}$ is characteristic of overdamped motion \cite{strogatz2015} while the scaling $\propto\epsilon^{-1/4}$ is characteristic of underdamped motion \cite{Gomez2017}. 

The key observation here is that most of the data in figure \ref{fig:slowingdown}  appears to be consistent with the same scaling law, namely $t_{\mathrm{PI}} \propto \epsilon^{-1/2}$ as $\epsilon \to 0$.  More precisely, we have fitted each data set (using least-squares) to a power law of the form $t_{\mathrm{PI}} = \alpha \epsilon^{-\beta}$ where $\beta > 0$; over all $27$ data sets considered in figure \ref{fig:slowingdown}, we find a mean value $\beta \approx 0.56$ with a standard deviation of $0.14$ in the fitted values. While an $\epsilon^{-1/2}$ scaling law has been identified as the source of slow dynamics in microbeam resonators \cite{zaitsev2012}, we believe this has not yet been properly appreciated as a generic feature of overdamped pull-in under DC loads. 

The slowing down observed in figure \ref{fig:slowingdown} motivates a more careful analysis of the dynamics of overdamped pull-in. In this paper we focus on the simplest possible electrostatic device: a parallel-plate actuator under a DC load. This single degree-of-freedom structure captures the balance between electrostatic and mechanical restoring forces that underlies the pull-in instability, without requiring details of the geometry of the device. It has been successfully used as a lumped-parameter model for more complex structures such as microbeams and microplates \cite{castaner1999}. Our analysis of the parallel-plate actuator therefore allows us to consider a generic MEMS device, upon taking appropriate values of the lumped parameters. Our central result is that the bottleneck behaviour observed near the pull-in transition is a type of saddle-node ghost, and so inherits the expected scaling law  \cite{strogatz2015}, with the pull-in time   $t_{\mathrm{PI}} \propto\epsilon^{-1/2}$ as $\epsilon \to 0$. While some data sets in figure \ref{fig:slowingdown} do not appear to follow this scaling, we suggest that the discrepancy is due to sensitivity to the precise value of the reported pull-in voltage, and propose a method to obtain a more accurate value based only on measured pull-in times.

The remainder of this paper is organized as follows. We begin in \S\ref{sec:formulation} by describing the equations governing the motion of the parallel-plate actuator. In \S\ref{sec:dynamics}, we solve the equations numerically when the system is perturbed just beyond the static pull-in transition. In the overdamped limit, we recover the bottleneck phenomenon reported previously \cite{rocha2004a}. We then perform a detailed asymptotic analysis of the solution structure in this regime, allowing us to derive an approximate expression for the pull-in time. In \S\ref{sec:experiments}, we compare our asymptotic prediction to the experimental and numerical data given in figure \ref{fig:slowingdown}. We show that the observed slowing down is well explained by our scaling law, and use our theory to collapse the data presented in figure \ref{fig:slowingdown} onto a master curve (see figure \ref{fig:datadimensionless}). Finally, in \S\ref{sec:conclusions}, we summarize and conclude our findings. 

\section{Theoretical formulation}
\label{sec:formulation}

\subsection{Governing equations}
We wish to understand the  bottleneck dynamics of a generic MEMS device, when the voltage is near the static pull-in transition. As the bottleneck is characterized by slow motions, and occurs well before the device comes into close contact with the actuating electrode \cite{rocha2004a}, we neglect compressibility and rarefaction effects in the squeeze film --- the fluid damping is assumed to be purely viscous \cite{veijola1995}. This is justified by numerical simulations \cite{missoffe2008} that show compressibility has very little effect on the pull-in time very close to the transition. Moreover,  we assume a constant damping coefficient, denoted by $b$, taken to be the effective value of the damping coefficient in the bottleneck.  Here $b$ is regarded as a lumped parameter that characterizes the properties of the squeeze film, including the thickness of the air gap, the ambient pressure, the fluid viscosity, and finite-width (border) effects, as well as any additional material damping that may be present. 

\begin{figure}
\centering
\includegraphics[width = 1\columnwidth]{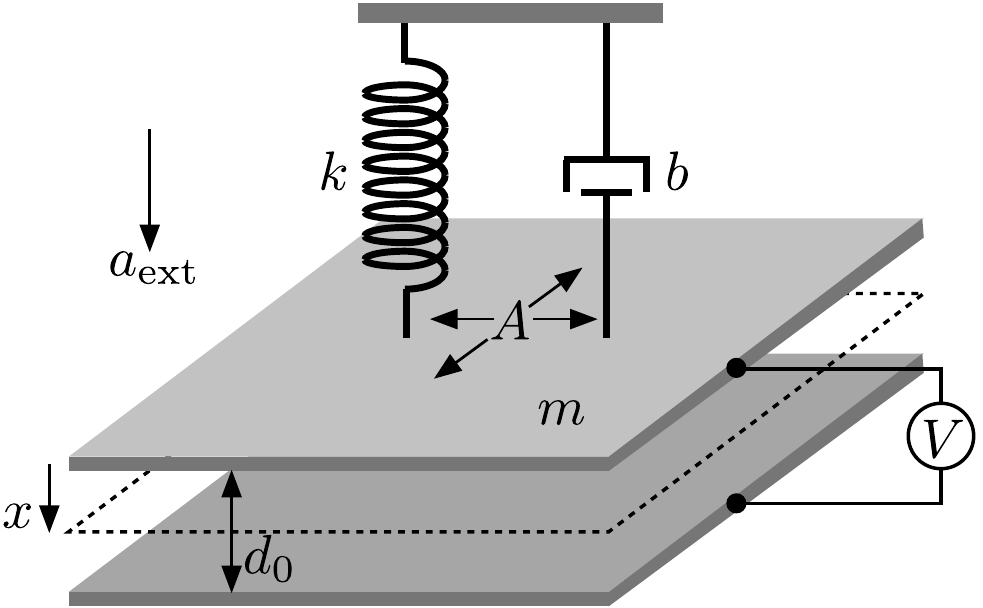} 
\caption{Schematic of the mass-spring parallel-plate capacitor. Fluid between the plates is represented by a linear dashpot of damping coefficient $b$.}
\label{fig:schematic}
\end{figure}

As the geometry of the device is also slowly varying when its motions are slow, the elastic restoring forces can be approximated to leading order as a linear spring with constant stiffness \cite{larose2010}. We denote the effective spring constant by $k$, which combines properties of the mechanical restoring force, such as the dimensions of the actuator, the material stiffness and any residual stress built into the elastic structure. In addition, we use a parallel-plate approximation of the electrostatic force; this is valid provided the aspect ratio of the air gap and the slopes of deformation are small \cite{batra2007}. In this way, our model becomes a single-degree-of-freedom mass-spring model \cite{pelesko2002}. Physically, it is equivalent to a parallel-plate capacitor, in which one plate is fixed while the other is attached to a linear spring and damper; see figure \ref{fig:schematic}.

The assumptions made above are not valid outside of the bottleneck phase, where the speed of the device is increased and the details of its geometry may become important. However, by choosing suitable values for the lumped parameters $b$ and $k$, we expect to correctly account for the length of the slow phase and hence approximate the total pull-in time, which is dominated by the time spent passing through the bottleneck. In particular, we discuss how a variable damping coefficient should be accounted for at the end of \S\ref{sec:dynamics} and in Appendix A.

As shown in figure \ref{fig:schematic}, the properties of the moving plate are its mass $m$, area $A$ and displacement $x$. The applied DC voltage is $V$, and $d_0$ is the gap thickness in the absence of any displacement ($x = 0$). We also account for an external acceleration $a_{\mathrm{ext}}$  of the whole device, which we assume is constant. This approximation is valid provided that  $a_{\mathrm{ext}}$ varies over a timescale much longer than the timescale of pull-in (typically $1-10~\mathrm{ms}$). 

Under these assumptions, the displacement of the moving plate, $x(t)$, obeys the equation of motion 
\beq
m \frac{\mathrm{d}^2 x}{\mathrm{d} t^2}+b \frac{\mathrm{d} x}{\mathrm{d} t} + k x = \frac{1}{2}\frac{\epsilon_0 A V^2}{(d_0 - x)^2}+m a_{\mathrm{ext}}. \label{eqn:odedim}
\eeq
Here the first term on the right-hand side is the electrostatic force in the parallel-plate approximation ($\epsilon_0$ is the permittivity of air), neglecting corrections due to fringing fields \cite{pelesko2002}. As initial conditions, we consider the case of a suddenly applied (step function) voltage with the plate initially at rest at the zero voltage state, i.e.~$x(0) = m a_{\mathrm{ext}}/k$ and $ \dot{x}(0) = 0$ (here and throughout $\dot{}$ denotes $\mathrm{d}/\mathrm{d}t$). These initial conditions are commonly used in applications of pull-in time in pressure sensors and accelerometers \cite{gupta1996,rocha2004a,dias2011,dias2015}. 

\subsection{Non-dimensionalization}
\label{sec:nondimensionalize}
To make the problem dimensionless, we note that a balance between viscous and spring forces in equation (\ref{eqn:odedim}) leads to the timescale $[t] = b/k$. It is natural to scale the displacement away from the zero voltage state with the maximum allowed displacement before contact occurs. This motivates introducing the dimensionless variables 
\beqn
T = \frac{t}{[t]}, \quad X = \frac{x - m a_{\mathrm{ext}}/k}{d_0 - m a_{\mathrm{ext}}/k}, \quad A_{\mathrm{ext}} = \frac{m a_{\mathrm{ext}}}{k d_0}.
\eeqn
Equation (\ref{eqn:odedim}) can then be written as 
\beq
Q^2 \frac{\mathrm{d}^2 X}{\mathrm{d} T^2} + \frac{\mathrm{d} X}{\mathrm{d} T} + X = \frac{\lambda}{(1 - X)^2},
 \label{eqn:odenondim}
\eeq
where $Q = \sqrt{m k}/b$ is the quality factor, and we have introduced the normalized voltage 
\beq
\lambda = \frac{1}{2}\frac{\epsilon_0 A  V^2}{k d_0^3 (1-A_{\mathrm{ext}})^3}.
\label{eqn:lambda}
\eeq
The initial conditions become
\beq
X(0) = \dot{X}(0) = 0,
\label{eqn:ICs}
\eeq and we note that, with this non-dimensionalization, contact between the electrodes occurs at $X = 1$, with physical solutions requiring $X < 1$.

The dimensionless parameter $\lambda$ is the key control parameter and may be interpreted as the ratio of the typical electrostatic force ($\sim \epsilon_0 A V^2/[2 d_0 ^2]$) to the spring force ($\sim k d_0$) \cite{pelesko2002}, together with an additional factor that depends on the acceleration of the device. For realistic MEMS devices we have $|A_{\mathrm{ext}}| \ll 1$, owing mainly to the small value of the mass $m$; for example, in the accelerometer designed by \cite{rocha2004a}, the range of accelerations encountered is $a_{\mathrm{ext}} \leq 80~\mathrm{mg}$, corresponding to $A_{\mathrm{ext}}  = O(10^{-3}) $ for their experimental parameters. We therefore consider only the case $\lambda \geq 0$ here. 

\begin{figure*}
\centering
\includegraphics[width =2\columnwidth]{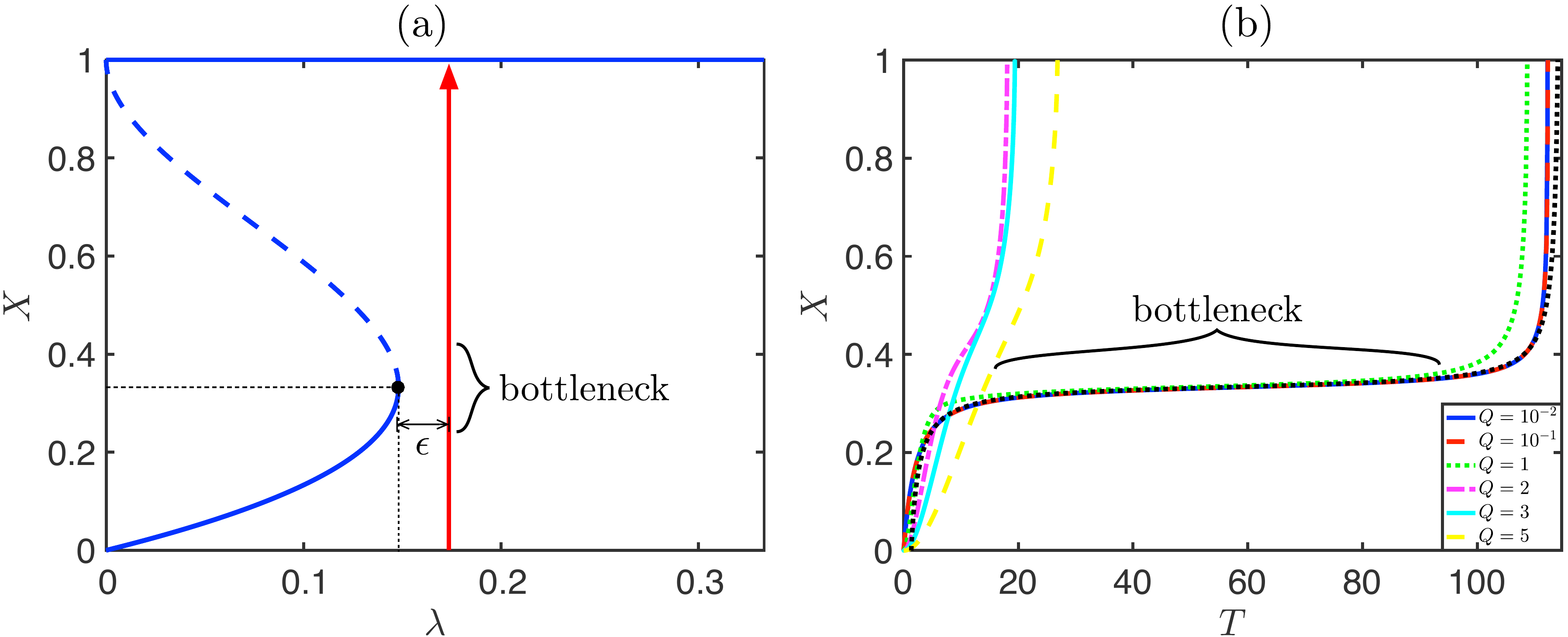} 
\caption{(a) Response diagram for the steady-state solutions of (\ref{eqn:odenondim}) (blue curves), which satisfy $X = \lambda/(1-X)^2$, as the dimensionless voltage $\lambda$ varies.  At $\lambda=4/27$ the stable equilibrium away from pull-in (lower solid curve) intersects an unstable solution (dashed curve) and disappears at a fold: a saddle-node bifurcation. A typical trajectory at fixed $\lambda$ beyond the pull-in transition is also shown (red arrow). (b) Dimensionless trajectories $X(T)$ satisfying (\ref{eqn:odenondim}) and  (\ref{eqn:ICs}) for $\epsilon = 10^{-3}$ and different quality factor $Q$ (coloured curves; see legend). For later comparison, the asymptotic trajectory predicted by (\ref{eqn:bottlenecksolndim}) is also shown (black dotted curve).}
\label{fig:response}
\end{figure*}

\subsection{Steady solutions}
The behaviour of the steady-state solutions of equation (\ref{eqn:odenondim}) is well-known (see \cite{pelesko2002}, for example). Here we merely summarize the main results. For $0 \leq \lambda < \lambda_{\mathrm{fold}} = 4/27$,  there are two real solutions with $0<X<1$, one of which is linearly stable and the other  linearly unstable. At  $\lambda = \lambda_{\mathrm{fold}}$, these two solutions coincide and disappear at a saddle-node (fold) bifurcation with $X = X_{\mathrm{fold}} = 1/3$. For $\lambda > \lambda_{\mathrm{fold}}$, no real solutions exist. This is illustrated by the response diagram shown in figure \ref{fig:response}a.

Under quasi-static conditions, the fold point corresponds to where pull-in is observed experimentally (no equilibrium solution away from contact exists for  $\lambda > \lambda_{\mathrm{fold}}$), giving the static pull-in voltage and pull-in displacement in terms of the external acceleration as
\beqn
V_{\mathrm{PI}} = \sqrt{\frac{8 k d_0^3 (1-A_{\mathrm{ext}})^3}{27 \epsilon_0 A}}, \quad x_{\mathrm{PI}} = \frac{d_0}{3}\left(1+2A_{\mathrm{ext}}\right).
\eeqn 
In the case of zero external acceleration, this reduces to the classic (static) pull-in voltage and pull-in displacement of a parallel-plate capacitor, widely reported in the literature \cite{zhang2014}; we label these as $V_{\mathrm{SPI}}$ and $x_{\mathrm{SPI}}$ respectively.

\section{Pull-in dynamics}
\label{sec:dynamics}
We now consider the case when the system is perturbed just beyond the static pull-in transition, i.e.~we set
\beqn
\lambda = \lambda_{\mathrm{fold}}(1+\epsilon),
\eeqn
where $0 < \epsilon \ll 1$ is a small parameter capturing the distance beyond the pull-in transition ($\epsilon$ is shown schematically in figure \ref{fig:response}a). Using the expression (\ref{eqn:lambda}) and the fact that 
\beqn
\lambda_{\mathrm{fold}} = \frac{1}{2}\frac{\epsilon_0 A  V_{\mathrm{SPI}}^2}{k d_0^3},
\eeqn
we may write $\epsilon$ as
\beq
\epsilon = \frac{\lambda}{\lambda_{\mathrm{fold}}}-1 =  \frac{\left(V/V_{\mathrm{SPI}}\right)^2}{\left(1-A_{\mathrm{ext}}\right)^3}-1. \label{eqn:epsilon}
\eeq
We see that for a fixed actuation voltage $V > V_{\mathrm{SPI}}$, the external acceleration changes the effective perturbation $\epsilon$, with $\epsilon$ increasing as $A_{\mathrm{ext}}$ increases. The result will be an associated change in the pull-in time. This is the basis on which pull-in time accelerometers operate: by repeatedly inducing pull-in and measuring the resulting pull-in times, the external acceleration can be determined after a suitable calibration is performed \cite{dias2011}. In practice, pull-in times can be measured extremely accurately and with low noise by sensing large changes in capacitance using a high frequency  clock. The sensitivity of the pull-in time to changes in $\epsilon$ is therefore the primary factor that limits the sensitivity of the accelerometer.

The key observation, first reported by \cite{rocha2004a}, is that for quality factors $Q$ smaller than unity (i.e.~overdamped devices) the motion of the plate is slowed in a bottleneck as it passes the static pull-in displacement, $X_{\mathrm{fold}} = 1/3$. This behaviour is confirmed in figure \ref{fig:response}b, which displays the dimensionless trajectories $X(T)$ during pull-in for different values of $Q$. We have obtained these trajectories by integrating equation (\ref{eqn:odenondim}) numerically with initial conditions (\ref{eqn:ICs}) in \textsc{matlab}. As the ODE is singular in the limit $Q \to 0$ (we lose the second-order derivative needed to satisfy the initial conditions), we use the \textsc{matlab} routine \texttt{ode15s}, which employs a stiff solver  to capture transients in which the inertia of the plate cannot be neglected. 

We see from figure \ref{fig:response}b that for $Q \ll 1$, the bottleneck phase dominates the transient dynamics, and hence the total time taken to pull-in. The phase  becomes highly dependent on the damping as $Q$ is increased past unity, with virtually no bottleneck present for $Q \geq 2$. The duration of the bottleneck is also sensitive to the perturbation $\epsilon$, and appears to increase without bound as $\epsilon \to 0$. We now perform a detailed asymptotic analysis of equation (\ref{eqn:odenondim}) in the limit $Q \ll 1$, showing that the bottleneck phenomenon is an instance of a saddle-node ghost \cite{strogatz2015} whose duration scales as $\epsilon^{-1/2}$ as $\epsilon \to 0$.

\subsection{Solution structure for $Q \ll 1$}
We begin by considering the different leading order balances the solution passes through during pull-in. This analysis will confirm that the  bottleneck phase does indeed dominate the pull-in dynamics, as expected from figure \ref{fig:response}b: the bottleneck duration is much longer than any other timescale in the problem, including any intervals for which plate inertia is important. This will enable us to approximate the total pull-in time based on the duration of the bottleneck alone. 

\subsubsection{Early times:}
At early times, the initial conditions (\ref{eqn:ICs}) imply that the displacement $X$ is small. Linearizing equation (\ref{eqn:odenondim}) then gives
\beqn
Q^2 \frac{\mathrm{d}^2 X}{\mathrm{d} T^2} + \frac{\mathrm{d} X}{\mathrm{d} T} + X \sim \lambda (1 + 2X).
\eeqn
The solution satisfying $X(0) = \dot{X}(0) = 0$ is
\beq
X = \frac{\lambda}{1-2\lambda}\left(1+\frac{\alpha_-}{\alpha_+-\alpha_-}e^{\alpha_+ T}-\frac{\alpha_+}{\alpha_+-\alpha_-}e^{\alpha_- T}\right), \label{eqn:smallXsoln}
\eeq
where 
\beqn
\alpha_{\pm} =  \frac{-1\pm \sqrt{1-4 Q^2 (1-2\lambda)}}{2 Q^2}.
\eeqn
When $Q \ll 1$, we expand to find
\beqn
\alpha_+ = -(1-2\lambda) + O(Q^2), \quad  \alpha_- = -\frac{1}{Q^2} + O(1).
\eeqn
This shows that inertia may only be neglected for $T \gg Q^2$, when $e^{\alpha_- T}$ is exponentially small and the leading order terms in (\ref{eqn:smallXsoln}) become independent of $Q$. In this case the solution simplifies to
\beqn
X = \frac{\lambda}{1-2\lambda}\left[1-e^{-(1-2\lambda)T}\right].
\eeqn
It follows that the terms we neglected in linearizing equation (\ref{eqn:odenondim}), of size $O(X^2)$, only remain small provided $T \ll 1$ (as $\lambda \approx \lambda_{\mathrm{fold}} = 4/27$ is order unity). As $T$ reaches $O(1)$, this solution therefore breaks down and a different leading order balance emerges.

\subsubsection{Later times, $T \gtrsim 1$:}
Using the previous solution to evaluate the size of terms for $T = O(1)$ yields the updated balance
\beq
\frac{\mathrm{d} X}{\mathrm{d} T} + X \sim \frac{\lambda}{(1-X)^2},
\label{eqn:odefirstorder}
\eeq
with inertia now negligible. This equation can be solved  to give the displacement implicitly in terms of time  (e.g.~\cite{gupta1997}):
\beq
T = \int_0^{X} \frac{(1-\xi)^2}{\lambda - \xi (1-\xi)^2}~\mathrm{d}\xi
\label{eqn:implicitsoln}
\eeq
(Here matching into $T \ll 1$ requires the constant of integration to be zero.)

This solution is not uniformly valid during pull-in: close to contact the electrostatic force will grow very large, leading to fast motions where inertia becomes important again. We can use equation (\ref{eqn:odefirstorder}) directly to determine when this first occurs. Differentiating, we obtain
\beqn
\frac{\mathrm{d}^2 X}{\mathrm{d} T^2} + \frac{\mathrm{d} X}{\mathrm{d} T}  \sim \frac{2 \lambda}{(1-X)^3} \frac{\mathrm{d} X}{\mathrm{d} T}.
\eeqn
The ratio of the neglected inertia term to the damping term can then be evaluated as
\beqn
\frac{Q^2\ddot{X}}{\dot{X}} \sim Q^2 \left[ \frac{2 \lambda}{(1-X)^3}-1\right].
\eeqn
Away from $X = 1$, the term in square brackets is $O(1)$ and so inertia is unimportant when $Q \ll 1$. This first breaks down when $X = 1-O(Q^{2/3})$, at which point we have $\dot{X} = O(Q^{-4/3})$ (using (\ref{eqn:odefirstorder})) and $Q^2 \ddot{X} = O(Q^{-4/3})$. Note that these updated scalings must hold close to the pull-in time, which we denote $T = T_{\mathrm{PI}}$, as $X$ is close to $1$. Setting $T = T_{\mathrm{PI}}-O(Q^{\gamma})$ and seeking a balance between these terms shows that $\gamma = 2$, i.e.~these scalings hold inside the interval $T = T_{\mathrm{PI}}-O(Q^2)$.

In summary, for $Q \ll 1$ we have shown that inertia of the plate remains negligible for
\beqn
Q^2 \ll T \ll T_{\mathrm{PI}}-O(Q^2), \quad Q^2 \ll X \ll 1 -O(Q^{2/3}).
\eeqn
In particular, we conclude that the dynamics are first order when $X$ passes the static pull-in displacement $X_{\mathrm{fold}} = 1/3$. Because $\lambda$ is close to its value at the fold, where the spring force exactly balances the electrostatic force, it follows that the difference between these two forces will be very small around $X_{\mathrm{fold}}$. This explains the previous observation that the bottleneck is a type of meta-stable interval characterized by a balance of forces \cite{rocha2004a}. In fact, when $X = X_{\mathrm{fold}}$ we have 
\beqn
\frac{\mathrm{d} X}{\mathrm{d} T}\bigg|_{X = X_{\mathrm{fold}}} = \left[\frac{\lambda}{(1-X)^2}
-X\right]\bigg|_{X = X_{\mathrm{fold}}} = \frac{\epsilon}{3}.
\eeqn
As the velocity is very small but non-zero in the bottleneck, the system appears to `feel' the attraction of the equilibrium point at the pull-in transition; however, this delay is purely a remnant of the saddle-node bifurcation, and relies on no extra physics in the system. 

Note that for larger quality factors, $Q = O(1)$, this conclusion is not valid: as the dynamics are no longer first order, a small net force does not imply slow dynamics. The inertia of the plate `carries' it through the bottleneck without significant slow down, as is evident from the trajectories in figure \ref{fig:response}b for $Q \geq 2$.  It can also be observed that the pull-in time does not simply decrease monotonically in this regime as $Q$ is increased (e.g.~the pull-in for $Q=2$ is faster  than that for $Q=5$ in figure \ref{fig:response}b): while high inertia carries the plate quickly through the bottleneck, it also slows down the initial dynamics, as the plate must be accelerated from its rest position.

\subsubsection{Bottleneck analysis:}
\label{sec:bottleneckanalysis}
We now consider the solution inside the bottleneck phase. While we can make progress using the implicit solution (\ref{eqn:implicitsoln}), we instead analyse equation (\ref{eqn:odenondim}) directly. The method we present is more general as it can be applied to systems for which no analytical solution is available. 

When the solution is close to the static pull-in displacement we have
\beqn
X = X_{\mathrm{fold}} \left[1+\tilde{X}(T)\right],
\eeqn
where $|\tilde{X}| \ll 1$. Using $\lambda = \lambda_{\mathrm{fold}}(1+\epsilon) $, the electrostatic force can then be expanded as
\beqn
\frac{\lambda}{(1-X)^2} = X_{\mathrm{fold}}\left[1+\epsilon+\tilde{X}+\frac{3}{4}\tilde{X}^2\right]+O(\epsilon \tilde{X},\tilde{X}^3).
\eeqn
Substituting into (\ref{eqn:odefirstorder}) and neglecting terms of $O(\epsilon \tilde{X},\tilde{X}^3)$, we obtain
\beq
\frac{\mathrm{d} \tilde{X}}{\mathrm{d} T}  \sim \epsilon + \frac{3}{4} \tilde{X}^2. 
\label{eqn:bottleneckeqn}
\eeq

Equation  (\ref{eqn:bottleneckeqn}) is valid in the regime $\epsilon \ll |\tilde{X}| \ll 1$, i.e.~the neglected terms of $O(\epsilon \tilde{X},\tilde{X}^3)$ are smaller than the retained terms. 
In particular, we note the importance of retaining the quadratic term. This term is neglected in the approach taken by \cite{rocha2004b}; an analysis of their solution shows that it incorrectly predicts the pull-in time scales as $\epsilon^{-1}$ as $\epsilon \to 0$.

Up to numerical pre-factors, equation (\ref{eqn:bottleneckeqn}) is the normal form for a saddle-node bifurcation \cite{strogatz2015}. This reflects the bifurcation structure underlying the pull-in transition: the first term on the right-hand side is the normalized perturbation to the bifurcation parameter (either due to a change in voltage or external acceleration), and the quadratic term is the nonlinearity that characterizes the bifurcation as being of saddle-node type (locally parabolic near the fold). In this way, equation (\ref{eqn:bottleneckeqn}) is generic for the dynamics of overdamped MEMS devices close to the pull-in transition. Similar evolution equations have been obtained  using a single degree-of-freedom approximation for a microbeam \cite{krylov2004}, and in a MEMS resonator modelled as a Duffing-like oscillator  \cite{zaitsev2012}. However, our approach here offers new insight into why this equation should apply more generally.

The solution of  (\ref{eqn:bottleneckeqn})  is
\beq
\tilde{X} \sim \frac{2}{3}\sqrt{3 \epsilon} \tan\left[\frac{1}{2}\sqrt{3 \epsilon} (T-T_0)\right], \label{eqn:bottlenecksoln}
\eeq
for some constant $T_0$.  In the immediate vicinity of the static pull-in displacement, where $|\tilde{X}| \ll \epsilon^{1/2}$, the term in square brackets in (\ref{eqn:bottlenecksoln}) is much smaller than unity. Here the solution simplifies to
\beqn
\tilde{X} \sim \epsilon (T-T_0),
\eeqn
so that the displacement evolves linearly in time in the middle of the bottleneck. Outside of this interval, the tangent function captures how the plate begins to accelerate away from the static pull-in displacement. We note that as this linear behaviour is precisely the solution of equation (\ref{eqn:bottleneckeqn}) upon neglecting the quadratic term in favour of the term in $\epsilon$, we deduce that (\ref{eqn:bottlenecksoln}) is asymptotically valid for all $|\tilde{X}|\ll 1$ (rather than just $\epsilon \ll |\tilde{X}| \ll 1$).

The solution (\ref{eqn:bottlenecksoln}) appears to undergo finite-time blow-up as the term in square brackets approaches $\pm \pi/2$. However, as soon as $\tilde{X}$ grows comparable to $O(1)$, our original assumption $|\tilde{X}| \ll 1$ is no longer valid and the solution breaks down. In terms of the diagram in figure \ref{fig:response}a, this means that the displacement has left the vicinity of the fold point and a local analysis can no longer be applied. Upon making use of the expansion $\tan x \sim\pm (\pi/2 \mp x)^{-1}$ as $x \to \pm \pi/2$, it follows that the solution accelerates according to the power law
\beqn
\tilde{X} \sim \frac{\pm 4/3}{\pi/\sqrt{3\epsilon}\mp(T-T_0)}.
\eeqn
Here the minus sign corresponds to initially entering the bottleneck ($\tilde{X} < 0$), while the plus sign corresponds to leaving the bottleneck towards pull-in ($\tilde{X} > 0$). We deduce that $\tilde{X} = O(1)$ when
\beqn
T-T_0 \sim \pm \frac{\pi}{\sqrt{3\epsilon}}+O(1).
\eeqn
The duration of the bottleneck, denoted $T_{\mathrm{bot}}$, is simply the difference between these two values and so we have $T_{\mathrm{bot}} = 2\pi/\sqrt{3\epsilon}+O(1)$.

The bottleneck dominates the time spent in the regime where the dynamics are first order. Moreover, we showed that inertia is only important in intervals of duration $O(Q^2)$ ($\ll 1$) around $T = 0$ and $T = T_{\mathrm{PI}}$. It follows that the total pull-in time is equal to the bottleneck time to leading-order:
\beq
T_{\mathrm{PI}} = \frac{2\pi}{\sqrt{3\epsilon}}+O(1).
\label{eqn:pullintimeND}
\eeq

\begin{figure*}
\centering
\includegraphics[width = 2\columnwidth]{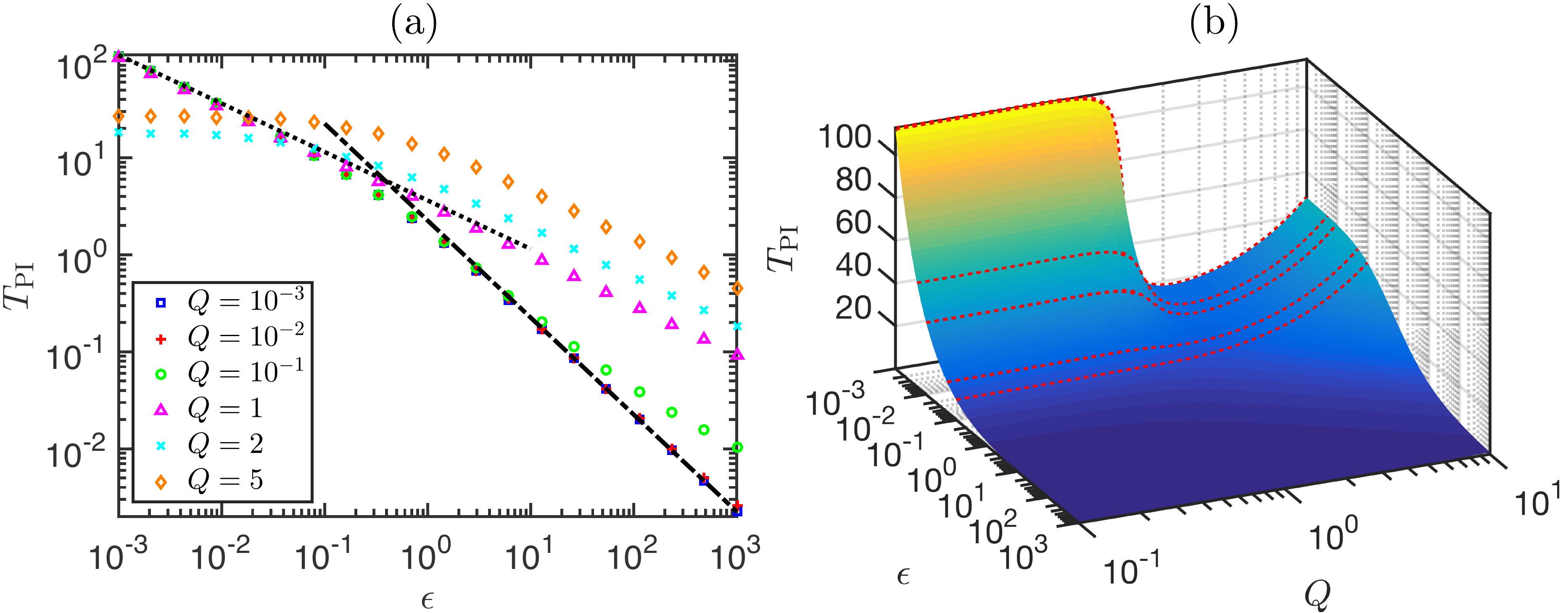} 
\caption{Pull-in times $T_{\mathrm{PI}}$ determined from the numerical solution of (\ref{eqn:odenondim}) with initial conditions (\ref{eqn:ICs}). (a) Numerical results for fixed $Q$ and variable $\epsilon$ (symbols, see legend), together with the asymptotic prediction $T_{\mathrm{PI}} \sim 2\pi/\sqrt{3\epsilon}$ valid for $\epsilon \ll 1$ and $Q \ll 1$ (dotted line), and the prediction $T_{\mathrm{PI}} \sim 9/(4\epsilon)$ valid for $\epsilon \gg 1$ and  $Q \ll 1$ (dashed-dotted line) \cite{gupta1997}. (b) Surface plot of $T_{\mathrm{PI}}$ as a function of $\epsilon$ and $Q$. Also shown are slices through the surface at values $\epsilon \in \lbrace 10^{-3},5\times 10^{-3},10^{-2},5\times 10^{-2},10^{-1} \rbrace$ (red dotted curves). }
\label{fig:pullintime}
\end{figure*}

To validate the prediction (\ref{eqn:pullintimeND}), we numerically determine the pull-in time by integrating the full ODE (\ref{eqn:odenondim}) with initial conditions (\ref{eqn:ICs}). As equation (\ref{eqn:odenondim}) is singular at $X = 1$, we use event location to stop integration as soon as $(1 - X) < \mathrm{tol}$ for some tolerance $\mathrm{tol}$, and the corresponding time at this point then gives the pull-in time. The accuracy of this method can be justified by analysing the behaviour of  (\ref{eqn:odenondim}) very close to pull-in, where a power law solution can be extracted; we use $\mathrm{tol} = 10^{-5}$, which guarantees an accuracy of $O(10^{-6})$ in the computed pull-in time when we restrict to $Q \leq 10$. In figure \ref{fig:pullintime}a we plot the computed times as a function of the normalized perturbation $\epsilon$. We conclude that the asymptotic prediction (\ref{eqn:pullintimeND}) approximates the pull-in time extremely well for moderately small quality factors $Q \lesssim 1$ and perturbations $\epsilon \lesssim 10^{-1}$.

Figure \ref{fig:pullintime}b shows a surface plot of the computed pull-in times for a range of values of $\epsilon$ and $Q$. As well as showing that the dynamics become very slow as $\epsilon \to 0$ with $Q$  fixed, we observe that, with $\epsilon$ fixed, the  dependence of the pull-in time on the quality factor $Q$ is non-monotonic. In particular, when we fix $\epsilon \lesssim 10^{-2}$, a minimum in  $T_{\mathrm{PI}}$ is obtained at $Q \approx 2$; within a narrow range of $Q$ close to this value, $T_{\mathrm{PI}}$ varies significantly. While this minimum may seem surprising at first, it is the result of inertia being small enough for the plate to be rapidly accelerated from its rest position but large enough that it passes the pull-in displacement without significant slowing down in a bottleneck. If we imagine fixing the actuation voltage $V$ near $V_{\mathrm{SPI}}$ and varying the plate mass $m$, so that $Q$ is varied while all other parameters are fixed, then this corresponds to a value of $m$ that minimizes the pull-in time. This may be relevant to switching applications where the pull-in time needs to be minimized without increasing the voltage significantly \cite{castaner1999} (since increasing the voltage would increase the total energy consumed). 

Currently, the constant $T_0$ appearing in the bottleneck solution (\ref{eqn:bottlenecksoln}) remains undetermined. This corresponds to the time at which $\tilde{X} = 0$ (when the displacement is equal to the static pull-in displacement $X_{\mathrm{fold}}$). However, we can find the value of $T_0$ by a symmetry argument. From the solution (\ref{eqn:bottlenecksoln}), we see that  the displacement about the static pull-in displacement is antisymmetric, i.e.~we have $\tilde{X} \to -\tilde{X}$ as $(T-T_0) \to -(T-T_0)$. (This is a consequence of the dynamics being first order, and the symmetry of the quadratic nonlinearity in equation (\ref{eqn:bottleneckeqn}).) As the bottleneck phase dominates the entire motion in the limit $\epsilon \ll 1$, it follows that, to leading order in $\epsilon$, the value of $T_0$ is simply half of the bottleneck time: $T_0 \sim \pi/\sqrt{3\epsilon}$. The rescaled displacement in the bottleneck, (\ref{eqn:bottlenecksoln}), can then be written as
\beqn
\tilde{X} \sim \frac{2  \sqrt{3\epsilon}}{3} \tan\left[\frac{\sqrt{3 \epsilon}}{2} T-\frac{\pi}{2}\right].
\eeqn
The unscaled displacement, $X$, then becomes
\beq
X \sim \frac{1}{3} + \frac{2  \sqrt{3\epsilon}}{9} \tan\left[\frac{\sqrt{3 \epsilon}}{2} T-\frac{\pi}{2}\right]. \label{eqn:bottlenecksolndim}
\eeq
This compares well to the trajectories obtained by numerical integration of the full system; see figure \ref{fig:response}b, where the analytical prediction is almost indistinguishable from numerical results with $Q \ll 1$.  As the motions are so fast outside the bottleneck, we see that (\ref{eqn:bottlenecksolndim}) also provides a good description of the global dynamics (restricting $X$ to the interval $[0,1]$), despite the fact that the assumptions made in deriving (\ref{eqn:bottlenecksolndim}) are only strictly valid in the bottleneck phase.

We note that some caution is needed when using a constant damping coefficient, as in our approach here: in reality the damping coefficient may itself depend on the current gap thickness. Indeed, simulations that use a constant damping coefficient corresponding to the initial gap thickness have been shown  to give large errors  \cite{rocha2004a}. However, using the damping coefficient  appropriate in the bottleneck phase of the motion correctly accounts for the duration of the bottleneck, and hence provides a good approximation of the total time taken to pull-in (see Appendix A).

\section{Data comparison}
\label{sec:experiments}
In \S\ref{sec:dynamics} we derived a scaling law for the slowing down of a parallel-plate actuator close to the pull-in transition. In dimensional form, this predicts that the pull-in time increases as
\beq
t_{\mathrm{PI}} \sim \frac{b}{k} \frac{2\pi}{\sqrt{3 \epsilon}} \quad \mathrm{where} \quad \epsilon =   \frac{\left(V/V_{\mathrm{SPI}}\right)^2}{\left(1-A_{\mathrm{ext}}\right)^3}-1.
\label{eqn:pullintime}
\eeq
This result is valid for $0 < \epsilon \ll 1$ and small quality factor, $Q \ll 1$. As discussed at the start of \S\ref{sec:formulation}, we expect that this result also describes the dynamics of a generic MEMS device operating in overdamped conditions; here we regard the damping coefficient $b$ and spring constant $k$ as lumped parameters that encapsulate the properties of the squeeze film and the mechanical restoring force during the bottleneck phase, respectively. 

We now compare our prediction to pull-in data reported in the literature, both from experiments and numerical simulations. The details of each data set are summarized in table \ref{table:parallelplate} for parallel-plate devices, and in table \ref{table:microbeam} for microbeam devices. These provide the relevant parameter values in each study, and the type of model used (for numerical simulations).  We have separated the data so that only the actuation voltage or the acceleration is varying within each data set, corresponding to a particular row in the tables. Where the properties of the actuator or the squeeze film have changed within a single reference, the data have therefore been separated into different rows of the tables. 

For  data on parallel-plate actuators (table \ref{table:parallelplate}), the relevant parameters are the ratio of the actuation voltage to the pull-in voltage $V/V_{\mathrm{SPI}}$, external acceleration $a_{\mathrm{ext}}$, pull-in voltage $V_{\mathrm{SPI}}$, initial gap thickness $d_0$, plate mass $m$, and the damping coefficient in the bottleneck phase, $b$.  For the data on microbeams (table \ref{table:microbeam}), the parameters are the pull-in voltage $V_{\mathrm{SPI}}$, initial gap thickness $d_0$, beam length $L$, beam thickness $h$, beam width $w$, and Young's modulus $E$. (In both tables, blank entries indicate that no value is provided in the reference.) A wide range of values are exhibited in these parameters across the studies. We also report any additional effects that may be present in experiments and simulations; these include residual stress, rarefaction effects, partial field screening, varying ambient pressures, and different boundary conditions for microbeams.

\begin{sidewaystable*}
\vspace{400pt}
\caption{Summary of the pull-in times of parallel-plate actuators reported in the literature. \label{table:parallelplate}}
\begin{indented}
\item[]
    \begin{tabular}{m{0.5cm}m{1.5cm}m{1cm}m{1cm}m{0.8cm}m{0.8cm} m{0.8cm}m{0.8cm}m{1.7cm}m{2cm}m{2cm}m{1.5cm}m{1.2cm}m{1cm}}
    \br
  Ref. &  Data type & $V/V_{\mathrm{SPI}}$ &  $a_{\mathrm{ext}}$ & $d_0$ & $m$ & $k$  & $V_{\mathrm{SPI}}$  & Reported $b$ & Model &  Additional & Fitted $b$ & Fitted $Q$ & Legend \\
   & & & & $ (\mu\mathrm{m})$ & $(\mu\mathrm{g})$ & $(\mathrm{N}\mathrm{m}^{-1})$  & $(\mathrm{V})$  &  $(\mathrm{g}\mathrm{s}^{-1})$ & & effects & $(\mathrm{g}\mathrm{s}^{-1})$ & & \\ 
   \mr
           \cite{gupta1996} & Simulation & Variable & 0 & 2.07 & 0.114 & 6.33 & 8.76 & 0.35 & MS, LD &  & 0.330 & 0.0813 & \includegraphics[scale = 1]{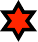}  \\     
           \cite{gupta1996} & Simulation & Variable & 0 & 2.07 & 0.132 & 2.94 &  5.54 & 0.35 &  MS, LD &  & 0.296 & 0.0668 &\includegraphics[scale = 1]{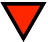}  \\     
             \cite{nijhuis1999}$^{\dagger}$ & Simulation & Variable & 0 & 25 & 1000 & 40 &  72.3 & & MS, CSQFD & SBC & 21.4 & 0.296 &\includegraphics[scale = 1]{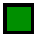} \\     
			\cite{rocha2004a} & Experiment & 1.0003 & Variable & 2.25 & 4.27 & 1.2930  & & 0.192 & N/A &  & 0.182 & 0.408 &  \includegraphics[scale = 1]{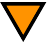} \\ 
			\cite{rocha2004a} & Experiment & 1.0005 & Variable & 2.25 & 4.27 & 1.2930  & & 0.192 & N/A &  &  0.182 & 0.408 & \includegraphics[scale = 1]{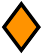} \\ 
			\cite{rocha2004a} & Experiment & 1.0009 & Variable & 2.25 & 4.27 & 1.2930  & & 0.192 & N/A &  &  0.182 & 0.408 & \includegraphics[scale = 1]{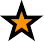} \\ 
			\cite{rocha2004a} & Simulation & 1.0003 & Variable & 2.25 & 4.27 & 1.2930  & & 0.192 & MS, CSQFD & SBC & 0.182 & 0.408 & \includegraphics[scale = 1]{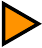} \\ 
			\cite{rocha2004a} & Simulation & 1.0005 & Variable & 2.25 & 4.27 & 1.2930  & & 0.192 & MS, CSQFD & SBC & 0.182  & 0.408 &\includegraphics[scale = 1]{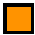} \\ 
			\cite{rocha2004a} & Simulation & 1.0009 & Variable & 2.25 & 4.27 & 1.2930 & & 0.192 & MS, CSQFD  & SBC & 0.182 & 0.408 & \includegraphics[scale = 1]{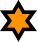} \\ 
			  \cite{dias2011} & Experiment & 1.01 & Variable & 2.25$^*$ & 249$^*$ & 3.33$^*$ & 2.931 & 2.88$^*$ & N/A & & 1.40 & 0.651 & \includegraphics[scale = 1]{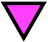}\\
			  \cite{dias2011} & Experiment & Variable & 0 & 2.25$^*$ & 249$^*$ & 3.33$^*$ & 2.931 & 2.88$^*$ & N/A & & 1.40 & 0.651 &  \includegraphics[scale = 1]{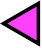}\\
			  \cite{dias2011} & Simulation & 1.01 & Variable & 2.25 & 249 & 3.33 & 2.916 & 2.88 & MS, CSQFD & SBC, BE & 1.40 & 0.651 & \includegraphics[scale = 1]{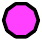} \\ 
			   \cite{dias2011} & Simulation & Variable & 0 & 2.25 & 249 & 3.33 & 2.916 & 2.88 & MS, CSQFD & SBC, BE & 1.40 & 0.651 & \includegraphics[scale = 1]{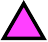}\\ 
	            \cite{dias2015} & Experiment & 1.05 & Variable & 3.00 & 497$^*$ & 2.40 & 2.94 & 4.80 & N/A & & 2.91 & 0.375 & \includegraphics[scale = 1]{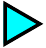}\\
	             \cite{dias2015} & Experiment & Variable & 0 & 3.00 & 497$^*$ & 2.40 & 2.94 & 4.80 & N/A & & 2.91 & 0.375 & \includegraphics[scale = 1]{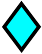} \\
	             \br
    \end{tabular}
\item[] MS, single degree-of-freedom mass-spring model; LD, linear damping with constant coefficient; CSQFD, compressible squeeze film damping (Reynolds equation); SBC, corrections due to slip boundary conditions (rarefaction effects); RS, residual stress; BE, border (finite-width) effects; N/A, not applicable 
  \item[] $^*$Experimental parameter reported as being obtained by modelling/estimated
\item[] $^{\dagger}$Electromagnetic force; data has been converted to equivalent DC voltage in electrostatic-electromagnetic analogy
\end{indented}
\end{sidewaystable*}

\begin{sidewaystable*}
\vspace{340pt}
\caption{Summary of the pull-in times of microbeam actuators reported in the literature. \label{table:microbeam}}
\begin{indented}
\item[]
    \begin{tabular}{m{0.5cm}m{1.5cm}m{0.8cm}m{0.8cm}m{0.8cm} m{0.8cm}m{0.8cm}m{0.8cm}m{3cm}m{3cm}m{1.5cm}m{1.5cm}}
       \br
  Ref. &  Data type & $d_0$ &  $L$ & $h$ & $w$ & $E$  & $V_{\mathrm{SPI}}$  & Model &  Additional effects & Fitted $[t]$ & Legend \\
   & &  $ (\mu\mathrm{m})$ &  $ (\mu\mathrm{m})$ & $ (\mu\mathrm{m})$ & $(\mu\mathrm{m})$ & $(\mathrm{GPa})$  & $(\mathrm{V})$  & &  & $(\mu\mathrm{s})$ & \\
   \mr
           \cite{gupta1996} & Experiment & 2.07 & 610 &  2.12 & 40  & 164 & 8.76 &  N/A &  &  52.1&\includegraphics[scale = 1]{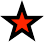} \\   
           \cite{gupta1996} & Experiment & 2.07 & 710 &  2.12 & 40  & 164 & 5.54 &  N/A &  & 101 &\includegraphics[scale = 1]{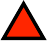} \\   
            \cite{gupta1996} & Simulation & 2.07 & 610 &  2.12 & 40  & 164 & 8.76 & PE, CC, CSQFD & RS & 52.1 &\includegraphics[scale = 1]{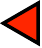} \\   
               \cite{gupta1996} & Simulation & 2.07 & 710 &  2.12 & 40  & 164 & 5.54 & PE, CC, CSQFD & RS & 101 &\includegraphics[scale = 1]{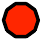} \\   
           \cite{gretillat1997} & Experiment &  & 300 &  &  & 200$^*$ & 48.7 &  N/A & FS & 0.444 &\includegraphics[scale = 1]{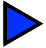} \\   
            \cite{gretillat1997} & Simulation &  & 300 &  &  & 200 & 48.7 & BE, CC, CSQFD & RS, FS & 0.444  &\includegraphics[scale = 1]{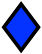} \\      
            \cite{gretillat1997} & Simulation &  & 300 &  &  & 200 & 48.5 & BE, SE, CSQFD & RS, FS & 0.444 &\includegraphics[scale = 1]{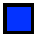} \\     
              \cite{hung1999} & Simulation & 2.3 & 610 & 2.2 & 40 & 149 & 8.76 & BE, CC, CSQFD & SBC, RS & 36.8 &\includegraphics[scale = 1]{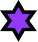} \\ 
               \cite{younis2003} & Simulation & 2.3 & 610 & 2.015 & 40 & 166 & 8.76 & BE, CC, LD & SS & 41.2 &\includegraphics[scale = 1]{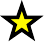} \\   
              \cite{missoffe2008} & Simulation & 2.3 & 610 & 2.2 & 40 & 149 & 8.76 & BE, CC, CSQFD & $0.1013~\mathrm{bar}^{\ddagger}$, SBC, RS & 9.38 &\includegraphics[scale = 1]{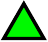} \\    
               \cite{missoffe2008} & Simulation & 2.3 & 610 & 2.2 & 40 & 149 & 8.76 & BE, CC, CSQFD & $1.013~\mathrm{bar}^{\ddagger}$, SBC, RS & 34.2 &\includegraphics[scale = 1]{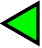} \\    
                \cite{missoffe2008} & Simulation & 2.3 & 610 & 2.2 & 70 & 149 & 8.76 & BE, CC, CSQFD & $0.1013~\mathrm{bar}^{\ddagger}$, SBC, RS & 21.9 &\includegraphics[scale = 1]{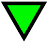} \\      	\br
    \end{tabular}
\item[] BE, dynamic beam equation; PE, dynamic (2D) plate equation; CC, clamped-clamped boundary conditions; SE, stepped-up boundary conditions; LD, linear damping with constant coefficient; CSQFD, compressible squeeze film damping (Reynolds equation); SBC, corrections due to slip boundary conditions (rarefaction effects); RS, residual stress; FS, partial field screening; SS, strain-stiffening; N/A, not applicable 
\item[] $^*$Experimental parameter reported as being obtained by modelling/estimated 
\item[] $^{\ddagger}$Ambient pressure
\end{indented}
\end{sidewaystable*}


The reported pull-in times were shown on logarithmic axes in figure \ref{fig:slowingdown} (as a function of the corresponding values of $\epsilon$). As well as generally confirming the expected scaling law that $t_{\mathrm{PI}} \propto\epsilon^{-1/2}$ as $\epsilon \to 0$, we see that this rescaling collapses data from accelerometers, where the acceleration is variable and the actuation voltage varies between each data set (all other parameters fixed); see the data of \cite{rocha2004a}, orange symbols. We also see a collapse in data over a single experimental system where the actuation mechanism changes, between varying the voltage (with zero acceleration) or varying the acceleration (with fixed voltage); see the data of \cite{dias2011}, magenta symbols. This verifies that $\epsilon$ is the correct dimensionless parameter to capture both variations in the voltage and external acceleration near the pull-in transition.


\begin{figure*}
\centering
\includegraphics[width = 2\columnwidth]{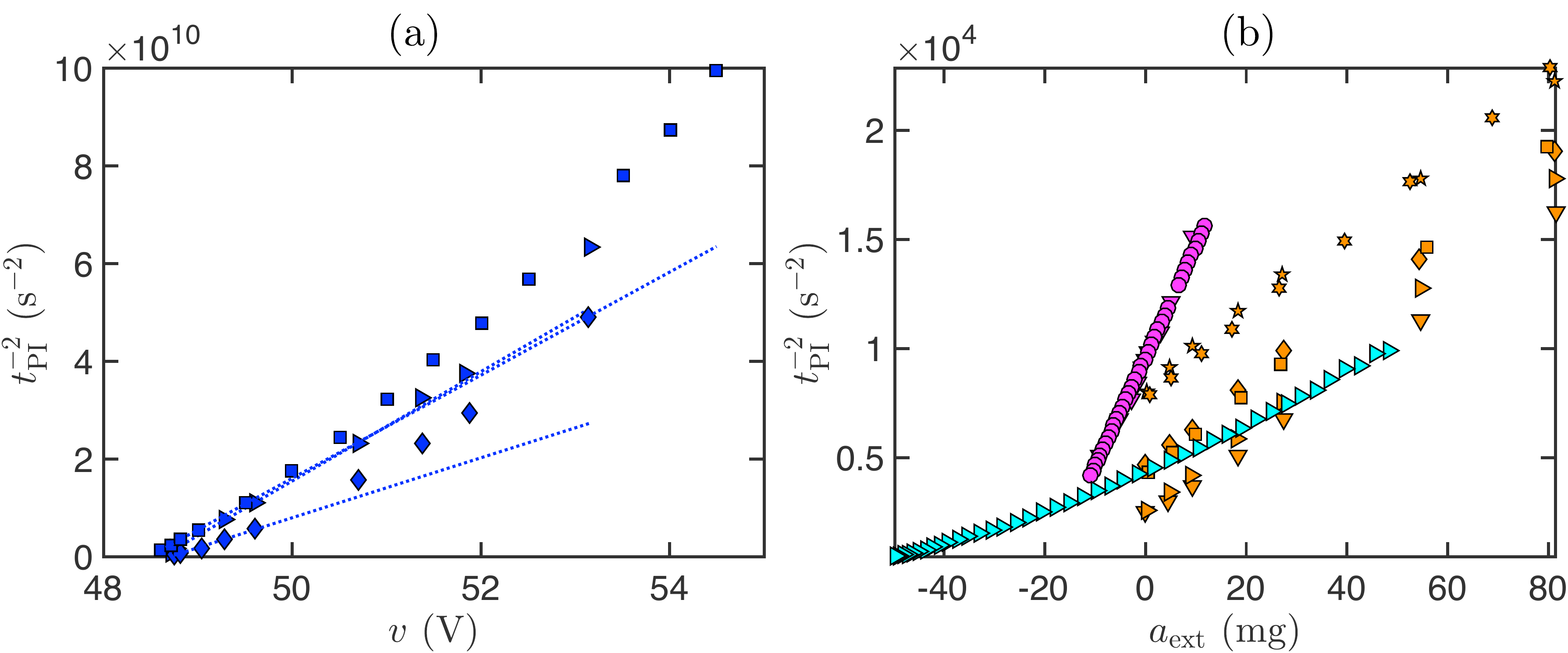} 
\caption{Re-scaling the pull-in times according to the scaling law $t_{\mathrm{PI}} = O(\epsilon^{-1/2})$ as $\epsilon \to 0$ for (a) the data of \cite{gretillat1997} and (b) all accelerometer data.}
\label{fig:datarescaled}
\end{figure*}

Some data sets plotted in figure \ref{fig:slowingdown} do not appear to follow the expected $\epsilon^{-1/2}$ scaling, curving downward slightly for small $\epsilon$. These include the experiments/simulations of \cite{gretillat1997} (blue symbols) and the experiments of \cite{dias2015} (cyan symbols). However, we propose that this is due to sensitivity to the reported value of the pull-in voltage: a small error introduces shifts in the computed values of $\epsilon$, which can cause large variations when plotted on logarithmic axes. Another way to test the scaling law, which eliminates this sensitivity, is to plot $t_{\mathrm{PI}}^{-2}$ as a function of voltage/external acceleration on linear axes. This is shown in figure \ref{fig:datarescaled}a for the data of \cite{gretillat1997} (blue symbols), and in figure \ref{fig:datarescaled}b for all accelerometer data. (Due to the large range of pull-in times under varying voltage,  figure \ref{fig:datarescaled}a shows only a subset of data, for clarity.)
In all cases a linear relationship is observed close to the pull-in transition, i.e.~as $t_{\mathrm{PI}}^{-2} \to 0$. 
A linear relationship here implies the expected  $\epsilon^{-1/2}$ scaling, because $\epsilon$ is linear in the voltage/acceleration close to the pull-in transition. For example, in the case of zero external acceleration, we have from (\ref{eqn:pullintime}):
\beqn
\epsilon = \left(\frac{V}{V_{\mathrm{SPI}}}\right)^2-1 \approx \frac{2}{V_{\mathrm{SPI}}}\left(V-V_{\mathrm{SPI}}\right),
\eeqn
when $V \approx V_{\mathrm{SPI}}$, and similarly in the case when the acceleration is varied. 

\subsection{Estimating the pull-in voltage}
The above analysis highlights the sensitivity of pull-in time to the actual pull-in voltage, which can be quite difficult to measure precisely --- for instance, quasi-statically increasing the voltage until pull-in occurs is subject to mechanical noise as well as imprecision in voltage measurements. There may also be rounding error in the reported pull-in voltage. Hence, in what follows we suggest an alternative approach. For the data of \cite{gretillat1997}, we have determined the best-fit (least-squares) line over the five data points that are closest to the pull-in transition (dotted lines on figure \ref{fig:datarescaled}a). By finding the intercept of each best-fit line with the horizontal axis, we are able to compute `corrected' values of the pull-in voltage. These are $\lbrace48.615,48.697,48.477\rbrace~\mathrm{V}$, which are in good agreement with the reported values of $\lbrace48.7,48.7,48.5\rbrace~\mathrm{V}$ respectively. This procedure may be applied more generally as a way to estimate the pull-in voltage based only on data for the pull-in times, rather than using the static behaviour of the system before pull-in occurs. 

\subsection{Estimating the pre-factor}
Finally, we show that it is possible to obtain good quantitative agreement with the predicted pre-factor in the scaling of (\ref{eqn:pullintime}), when we use realistic values of the lumped parameters $b$ and $k$. We make the pull-in times shown in figure \ref{fig:slowingdown} dimensionless using the timescale $[t] = b/k$. We can then use $[t]$ as a single fitting parameter to fit each data set to the dimensionless prediction $T_{\mathrm{PI}} \sim 2\pi/\sqrt{3\epsilon}$. This is consistent with the way we have separated each data set: as the properties of the squeeze film (e.g.~plate area, ambient pressure) and the mechanical restoring force (e.g~material stiffness, beam length) entering $b$ and $k$ do not vary in each data set, the timescale $[t]$ is fixed. Many of the references in tables \ref{table:parallelplate}--\ref{table:microbeam} have multiple data sets with the same timescale $[t]$; for example, when the actuation mechanism changes over a single experimental system (e.g.~the data of \cite{dias2011}, magenta symbols), and when numerical simulations use the same parameter values as experiments (e.g.~the data of \cite{rocha2004a}, orange symbols). In these cases we fit $[t]$ to only one set of experimental data and use this value to non-dimensionalize all of the data sets. 

\begin{figure*}
\centering
\includegraphics[width = 1.4\columnwidth]{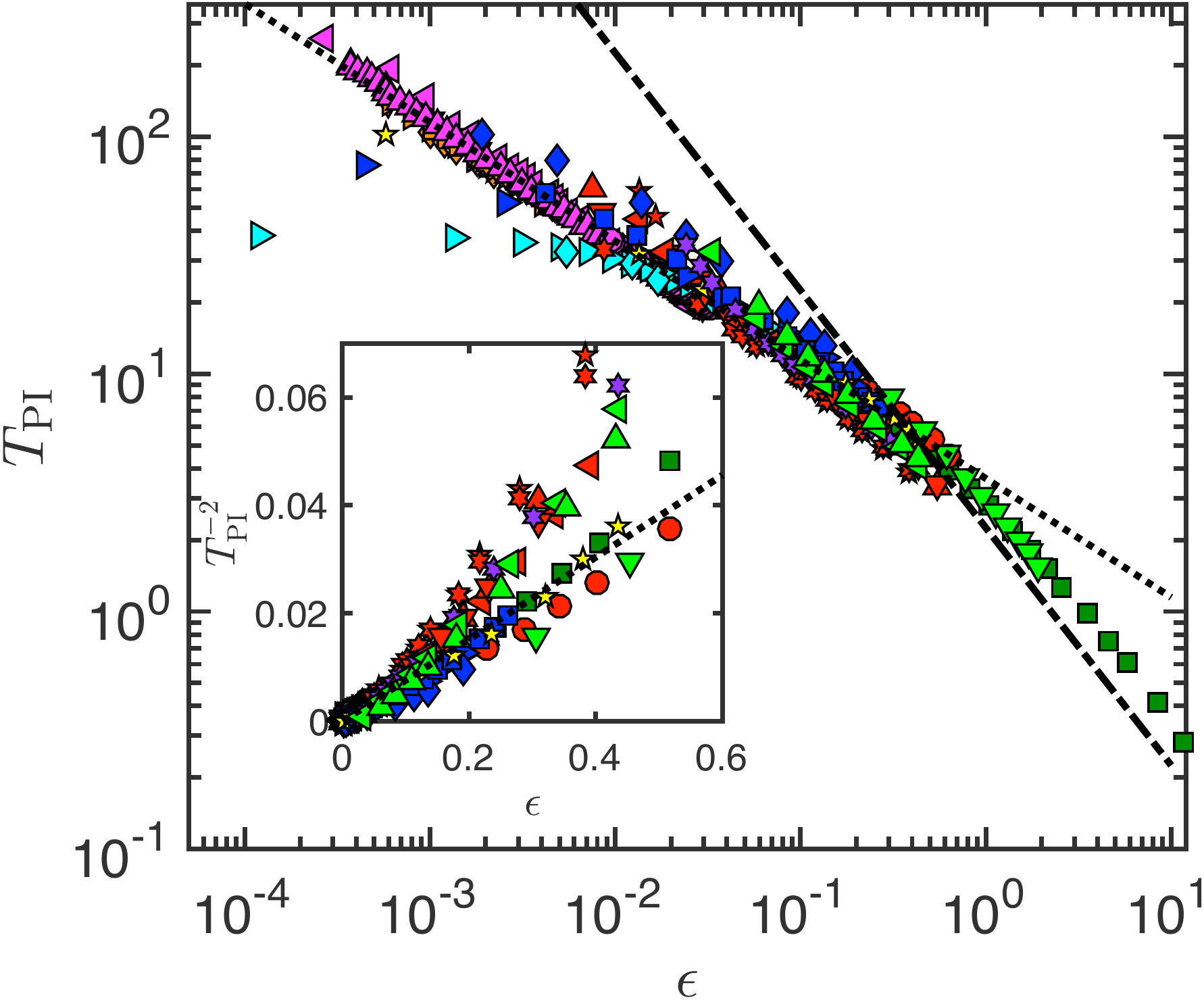} 
\caption{Main plot: Dimensionless pull-in times obtained by fitting the overdamped timescale $[t] = b/k$. Plotted for comparison is the prediction $T_{\mathrm{PI}} \sim 2\pi/\sqrt{3\epsilon}$ valid for $\epsilon \ll 1$ (black dotted line), as well as the large $\epsilon$ prediction $T_{\mathrm{PI}} \sim 9/(4\epsilon)$ (black dashed-dotted line) \cite{gupta1997}. Inset: The same data, re-scaled according to the scaling law $T_{\mathrm{PI}} = O(\epsilon^{-1/2})$ as $\epsilon \to 0$.}
\label{fig:datadimensionless}
\end{figure*}

The best-fit (least-squares) timescales $[t]$ are given in table \ref{table:microbeam} for the data on microbeams. For the data on parallel-plate actuators, the spring constant $k$ is usually a known design parameter, and so we use the reported values of $k$ to give the corresponding best-fit damping coefficient $b$. These compare reasonably well to approximate values obtained in numerical simulations of squeeze film damping (table \ref{table:parallelplate}). The corresponding quality factors $Q = \sqrt{m k}/b$ are all smaller than unity, so that the fitting performed here is consistent with our assumption that the devices are overdamped.

With the fitted values of $[t]$, we obtain excellent collapse over all data sets considered, up to the sensitivity to the value of the pull-in voltage used; see the main panel of figure \ref{fig:datadimensionless}. In the inset of figure \ref{fig:datadimensionless} we plot $T_{\mathrm{PI}}^{-2}$ as a function of $\epsilon$ on linear axes, which demonstrates the collapse for small $\epsilon$ without this sensitivity. 

\section{Conclusions}
\label{sec:conclusions}
In this paper we have considered the pull-in dynamics of overdamped MEMS devices. When the system is near the static pull-in voltage/acceleration, the motion is known to slow down considerably during a meta-stable or bottleneck phase. By considering a lumped-parameter parallel-plate model, we have shown that the bottleneck behaviour is an instance of a saddle-node ghost  \cite{strogatz2015}; the duration of the bottleneck increases $\propto\epsilon^{-1/2}$, where $\epsilon$ is the normalized distance of the system beyond the pull-in transition. A detailed asymptotic analysis then allowed us to evaluate the prefactor in this scaling law. The result is a simple analytical prediction for the total pull-in time: $t_{\mathrm{PI}} \sim (b/k) 2\pi/\sqrt{3 \epsilon}$, in which $b$ is the effective damping coefficient and $k$ is the lumped mechanical stiffness applicable to the bottleneck phase. This result complements previous studies that have calculated a similar asymptotic pull-in time for underdamped devices \cite{leus2008,joglekar2011}. 

The $\epsilon^{-1/2}$ scaling law explains the high sensitivity of the pull-in time observed in previous experiments and numerical simulations. Moreover, because the bottleneck phase dominates the dynamics during pull-in, the resulting pull-in time is relatively insensitive to what happens outside of the bottleneck region; this includes the precise geometry of the device, the effects of compressibility and air rarefaction, and the way in which stoppers limit the displacement before contact occurs. The implication is that a simple parallel-plate model, using lumped values for the damping coefficient and spring constant, is an effective means of capturing the behaviour of a complex MEMS device. Indeed, the wide range of available data collapsed onto a single master curve (figure \ref{fig:datadimensionless}), despite the number of additional effects that are present in the range of experiments and simulations analysed (summarized in tables \ref{table:parallelplate}--\ref{table:microbeam}). Moreover, while the assumption of a constant damping coefficient is often stated to give large errors \cite{hung1999,rocha2004a}, we have shown that, in the bottleneck regime, this assumption is sufficient to correctly predict the pull-in time.

The sensitivity of the bottleneck to external perturbations is the basis of using pull-in time as a sensing mechanism, as in some pressure sensors \cite{gupta1996,gupta1997} and accelerometers \cite{rocha2004a,dias2011,dias2015}. Currently, the lack of linearity in the response is considered to be the main disadvantage of these devices, and it has been suggested that the pull-in time curve might be linearized by the  introduction of extra forces \cite{dias2011}. Our expression for the pull-in time partly resolves the issue, as it provides a simple power law that can be used to calibrate a device. In addition, our introduction of the dimensionless parameter $\epsilon$, equation (\ref{eqn:epsilon}), captures both variations in the voltage and external acceleration near the pull-in transition. When plotted in terms of this parameter, we observe a collapse of data over experiments where either the voltage or the acceleration was varied. 

Finally, we discuss the conditions under which our analysis holds. We have considered only devices with low quality factors, so that inertia of the moving electrode can be neglected during the bottleneck phase. We also focussed on DC loads that are stepped from zero, since this loading type is commonly used in applications of the pull-in time. Nevertheless, our analysis may be adapted to other types of loading, provided the behaviour before pull-in remains quasi-static; for example, if the voltage is instead stepped from a positive value. However, in the case of a voltage sweep (e.g.~triangular wave), the quasi-static condition is not met and the  $\epsilon^{-1/2}$ scaling law will not apply.  Similarly, extremely close to the pull-in transition, mechanical noise will eventually become important and limit the system response. Nevertheless, we hope that the unifying perspective we have presented here will lead to new insights in the application of dynamic pull-in instabilities. 

\paragraph{Acknowledgments}

The research leading to these results has received funding from the European Research Council under the European Union's Horizon 2020 Programme/ERC Grant No. 637334 (D.V.) and the EPSRC Grant No. EP/ M50659X/1 (M.G.). The data that supports the plots within this paper and other findings of this study are available from http://dx.doi.org/10.5287/bodleian:7J84mXZ78. 

\section*{Appendix A: Assumption of a constant damping coefficient}
The assumption of a constant damping coefficient has often been reported to give large errors compared to simulations that incorporate a variable damping coefficient \cite{rocha2004a,hung1999}; based on this, it is argued that a variable damping coefficient should always be used when predicting the pull-in time for MEMS applications. For example, ref.~\cite{rocha2004a} consider the pull-in dynamics of a parallel-plate actuator, showing that a constant damping coefficient approximation leads to errors of up to $40\%$.  However, \cite{rocha2004a} use the value of the damping coefficient when the plate is in the zero voltage state, $b_{\mathrm{init}}$. This damping is much smaller than the value when the plate is near the static pull-in displacement,  $b_{\mathrm{PI}}$ (where the thickness of the air gap is around $2/3$ of the zero voltage thickness). Because the pull-in timescale $[t]$ depends linearly on the damping coefficient (for overdamped devices), and the system spends most of its time close to the pull-in displacement during the bottleneck phase, using  $b_{\mathrm{init}}$ will  lead to a significant under-prediction of the pull-in time. Here, we show that using $b_{\mathrm{PI}}$ (our approach in the main text) is sufficient to correctly predict the pull-in time.

We modify our spring-mass model to consider a variable damping coefficient $b(x)$:
\beq
m \frac{\mathrm{d}^2 x}{\mathrm{d} t^2}+b(x) \frac{\mathrm{d} x}{\mathrm{d} t} + k x = \frac{1}{2}\frac{\epsilon_0 A V^2}{(d_0 - x)^2}+m a_{\mathrm{ext}}. \label{eqn:odedimvaryingb}
\eeq
Ignoring compressibility and rarefaction effects, the incompressible Reynolds equation may be solved approximately in the parallel-plate geometry to give \cite{veijola1995}
\beqn
b(x) = \frac{\mu C}{(d_0-x)^3},
\eeqn
where $\mu$ is the air viscosity and $C$ is a constant that depends on the dimensions of the moving plate. The damping coefficient  corresponding to the zero voltage state, $x = m a_{\mathrm{ext}}/k$, is then
\beqn
b_{\mathrm{init}} = \frac{\mu C}{d_0^3 (1-A_{\mathrm{ext}})^3}.
\eeqn
Since a variable damping coefficient does not change the steady solutions, the static pull-in displacement is $x_{\mathrm{PI}} = (d_0/3)\left(1+2A_{\mathrm{ext}}\right)$, as before. The damping coefficient during the bottleneck phase is then
\beqn
b_{\mathrm{PI}} = \frac{27}{8}\frac{\mu C}{d_0^3 (1-A_{\mathrm{ext}})^3},
\eeqn
so that 
\beqn
\frac{b_{\mathrm{init}}}{b_{\mathrm{PI}}} = \frac{8}{27}.
\eeqn

We non-dimensionalize in a similar way to \S\ref{sec:nondimensionalize}, though now we set
\beqn
T = \frac{t}{b_{\mathrm{init}}/k}, \quad Q = \frac{\sqrt{m k}}{b_{\mathrm{init}}}.
\eeqn
Equation (\ref{eqn:odedimvaryingb}) then becomes
\beq
Q^2 \frac{\mathrm{d}^2 X}{\mathrm{d} T^2} + \frac{1}{(1-X)^3}\frac{\mathrm{d} X}{\mathrm{d} T} + X = \frac{\lambda}{(1 - X)^2},
 \label{eqn:odenondimvaryingb}
\eeq
and the initial conditions remain $X(0) = \dot{X}(0) = 0$. We may then perform a local analysis of equation (\ref{eqn:odenondimvaryingb}) when the solution is in the bottleneck phase, along similar lines to \S\ref{sec:bottleneckanalysis} (where now we Taylor expand the $(1-X)^3$ term about $X = X_{\mathrm{fold}} = 1/3$). This shows that the dimensional pull-in time, $t_{\mathrm{PI}}$, is given by
\beqn
t_{\mathrm{PI}} \sim \frac{b_{\mathrm{PI}}}{k} \frac{2\pi}{\sqrt{3 \epsilon}},
\eeqn
valid for $0 < \epsilon \ll 1$. We conclude that setting $b = b_{\mathrm{PI}}$ in our constant damping model (as done in the main text) yields the correct asymptotic expression for the pull-in time  (see equation \ref{eqn:pullintime}), while setting $b = b_{\mathrm{init}}$ will lead to a prediction of  $t_{\mathrm{PI}}$  that is a fraction $8/27$ (approximately $30\%$) of the true value. 

\begin{figure}
\centering
\includegraphics[width = 1\columnwidth]{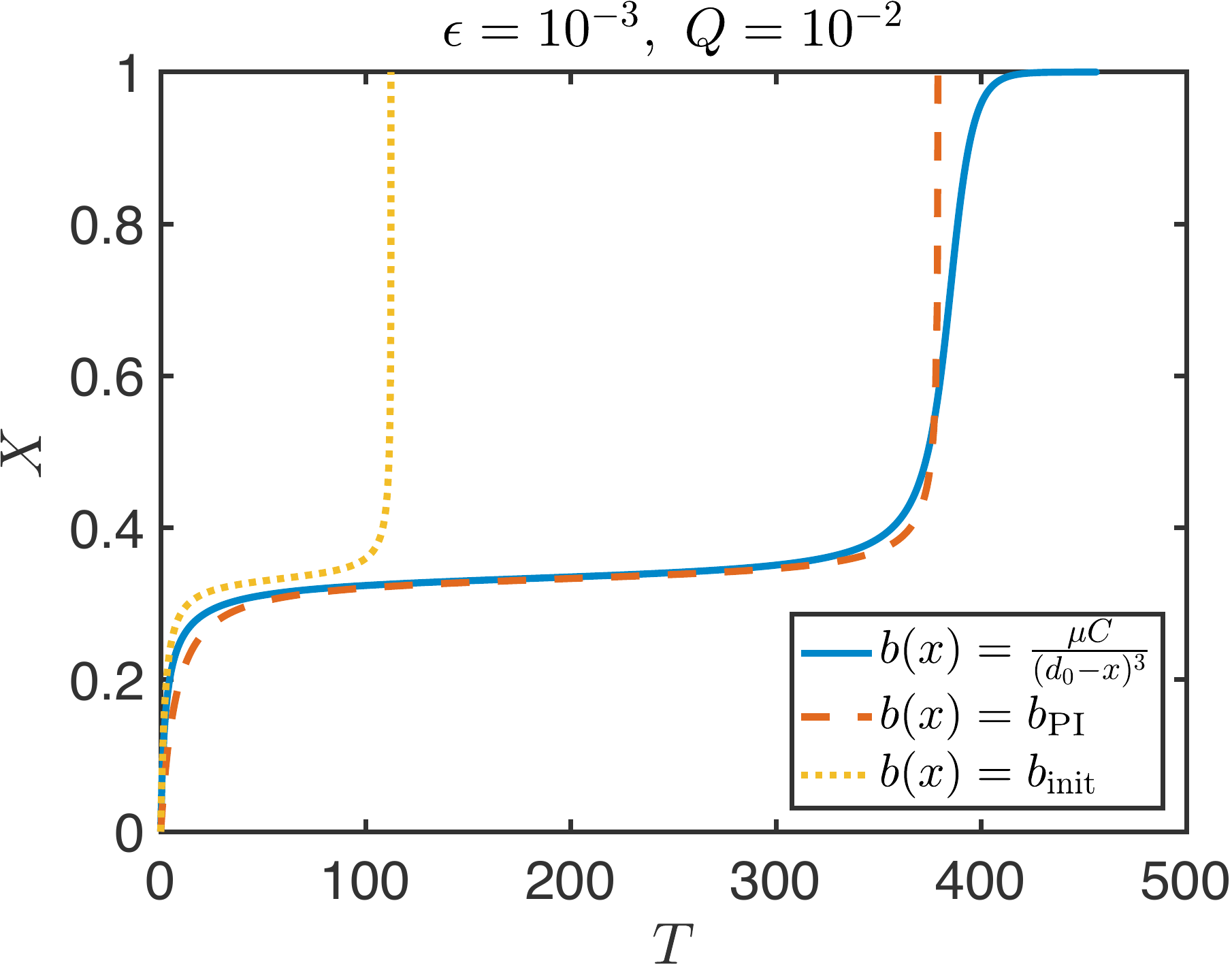} 
\caption{Trajectories obtained by numerical integration of equation (\ref{eqn:odedimvaryingb}) with different damping models $b(x)$ (see legend). Here dimensionless quantities are defined as in \S\ref{sec:nondimensionalize}, but now we set $T = t/(b_{\mathrm{init}}/k)$ and $Q = \sqrt{mk}/b_{\mathrm{init}}$.}
\label{fig:comparedamping}
\end{figure}

This is illustrated in figure \ref{fig:comparedamping}, which compares the numerical solution of the full equation (\ref{eqn:odenondimvaryingb}) with two approximate approaches: (i) the solution in which we instead assume a constant damping coefficient $b(x) = b_{\mathrm{PI}}$ (corresponding to setting $X = X_{\mathrm{fold}}$ in the $(1-X)^3$ term) and (ii) the solution with a constant coefficient $b(x) = b_{\mathrm{init}}$ (setting $X = 0$ in the $(1-X)^3$ term). We see that the constant coefficient $b_{\mathrm{PI}}$ successfully captures the duration of the bottleneck phase, and hence the total time taken to pull-in, while using $b_{\mathrm{init}}$  leads to large errors.

In ref.~\cite{rocha2004a} it is reported that using a constant coefficient $b_{\mathrm{init}}$ yields a pull-in time that is $60\%$  of that obtained using a variable damping coefficient. This is larger than the $\approx 30\%$ that we predict here. However, the simulations reported in \cite{rocha2004a} also incorporate compressibility and rarefaction effects in the squeeze film, which may account for this discrepancy.


\section*{References}
\begin{thebibliography}{10}

\bibitem{batra2007}
R.~C. Batra, M.~Porfiri, and D.~Spinello.
\newblock Review of modeling electrostatically actuated microelectromechanical
  systems.
\newblock {\em Smart Mater. Struct.}, 16(6):R23--R31, 2007.

\bibitem{krylov2008}
S.~Krylov, B.~R. Ilic, D.~Schreiber, S.~Seretensky, and H.~Craighead.
\newblock The pull-in behavior of electrostatically actuated bistable
  microstructures.
\newblock {\em J. Micromech. Microeng.}, 18(5):055026, 2008.

\bibitem{pelesko2002}
J.~A. Pelesko and D.~H. Bernstein.
\newblock {\em Modeling MEMS and NEMS}.
\newblock CRC press, 2002.

\bibitem{zhang2014}
W.-M. Zhang, H.~Yan, Z.-K. Peng, and G.~Meng.
\newblock Electrostatic pull-in instability in {MEMS/NEMS}: A review.
\newblock {\em Sens. Actuators A Phys}, 214:187--218, 2014.

\bibitem{younis2009}
M.~I. Younis and F.~Alsaleem.
\newblock Exploration of new concepts for mass detection in
  electrostatically-actuated structures based on nonlinear phenomena.
\newblock {\em J. Comput. Nonlin. Dyn.}, 4(2):021010, 2009.

\bibitem{osterberg1997}
P.~M. Osterberg and S.~D. Senturia.
\newblock {M-TEST}: a test chip for {MEMS} material property measurement using
  electrostatically actuated test structures.
\newblock {\em J. Microelectromech. Syst.}, 6(2):107--118, 1997.

\bibitem{desai2012}
A.~V. Desai, J.~D. Tice, C.~A. Apblett, and P.~J.~A. Kenis.
\newblock Design considerations for electrostatic microvalves with applications
  in poly (dimethylsiloxane)-based microfluidics.
\newblock {\em Lab Chip}, 12(6):1078--1088, 2012.

\bibitem{nguyen1998}
C.~T.-C. Nguyen, L.~P.~B. Katehi, and G.~M. Rebeiz.
\newblock Micromachined devices for wireless communications.
\newblock {\em Proc. IEEE}, 86(8):1756--1768, 1998.

\bibitem{larose2010}
R.~P. LaRose and K.~D. Murphy.
\newblock Impact dynamics of {MEMS} switches.
\newblock {\em Nonlinear Dyn.}, 60(3):327--339, 2010.

\bibitem{deng2017}
P.~Deng, N.~Wang, F.~Cai, and L.~Chen.
\newblock A high-force and high isolation metal-contact {RF MEMS} switch.
\newblock {\em Microsyst. Technol.}, 23(10):4699--4708, 2017.

\bibitem{gupta1997}
R.~K. Gupta and S.~D. Senturia.
\newblock Pull-in time dynamics as a measure of absolute pressure.
\newblock In {\em Proc. IEEE. Int. Workshop on MEMS (Nagoya, Jan. 1997)}, pages
  290--294. IEEE, 1997.

\bibitem{rocha2004a}
L.~A. Rocha, E.~Cretu, and R.~F. Wolffenbuttel.
\newblock Behavioural analysis of the pull-in dynamic transition.
\newblock {\em J. Micromech. Microeng.}, 14(9):S37, 2004.

\bibitem{dias2011}
R.~A. Dias, E.~Cretu, R.~Wolffenbuttel, and L.~A. Rocha.
\newblock Pull-in-based $\mu$g-resolution accelerometer: Characterization and
  noise analysis.
\newblock {\em Sens. Actuators A Phys}, 172(1):47--53, 2011.

\bibitem{dias2015}
R.~A. Dias, F.~S. Alves, M.~Costa, H.~Fonseca, J.~Cabral, J.~Gaspar, and L.~A.
  Rocha.
\newblock Real-time operation and characterization of a high-performance
  time-based accelerometer.
\newblock {\em J. Microelectromech. Syst.}, 24(6):1703--1711, 2015.

\bibitem{missoffe2008}
A.~Missoffe, J.~Juillard, and D.~Aubry.
\newblock A reduced-order model of squeeze-film damping for deformable
  micromechanical structures including large displacement effects.
\newblock {\em J. Micromech. Microeng.}, 18(3):035042, 2008.

\bibitem{nayfeh2005}
A.~H. Nayfeh, M.~I. Younis, and E.~M. Abdel-Rahman.
\newblock Reduced-order models for {MEMS} applications.
\newblock {\em Nonlinear Dyn.}, 41(1):211--236, 2005.

\bibitem{gretillat1997}
M.-A. Gr\'{e}tillat, Y.-J. Yang, E.~S. Hung, V.~Rabinovich, G.~K.
  Ananthasuresh, N.~F. De~Rooij, and S.~D. Senturia.
\newblock Nonlinear electromechanical behaviour of an electrostatic microrelay.
\newblock In {\em Proc. Int. Conf. on Solid State Sensors and Actuators
  (Transducers 1997, Chicago, IL)}, volume~2, pages 1141--1144. IEEE, 1997.

\bibitem{hung1999}
E.~S. Hung and S.~D. Senturia.
\newblock Generating efficient dynamical models for microelectromechanical
  systems from a few finite-element simulation runs.
\newblock {\em J. Microelectromech. Syst.}, 8(3):280--289, 1999.

\bibitem{younis2003}
M.~I. Younis, E.~M. Abdel-Rahman, and A.~Nayfeh.
\newblock A reduced-order model for electrically actuated microbeam-based
  {MEMS}.
\newblock {\em J. Microelectromech. Syst.}, 12(5):672--680, 2003.

\bibitem{nayfeh2007}
A.~H. Nayfeh, M.~I. Younis, and E.~M. Abdel-Rahman.
\newblock Dynamic pull-in phenomenon in {MEMS} resonators.
\newblock {\em Nonlinear Dyn.}, 48(1-2):153--163, 2007.

\bibitem{zaitsev2012}
S.~Zaitsev, O.~Shtempluck, E.~Buks, and O.~Gottlieb.
\newblock Nonlinear damping in a micromechanical oscillator.
\newblock {\em Nonlinear Dyn.}, 67(1):859--883, 2012.

\bibitem{chaterjee2009}
S.~Chaterjee and G.~Pohit.
\newblock A large deflection model for the pull-in analysis of
  electrostatically actuated microcantilever beams.
\newblock {\em J. Sound Vib.}, 322(4):969--986, 2009.

\bibitem{mccarthy2002}
B.~McCarthy, G.~G. Adams, N.~E. McGruer, and D.~Potter.
\newblock A dynamic model, including contact bounce, of an electrostatically
  actuated microswitch.
\newblock {\em J. Microelectromech. Syst.}, 11(3):276--283, 2002.

\bibitem{das2009}
K.~Das and R.~C. Batra.
\newblock Pull-in and snap-through instabilities in transient deformations of
  microelectromechanical systems.
\newblock {\em J. Micromech. Microeng.}, 19(3):035008, 2009.

\bibitem{krylov2010}
S.~Krylov and N.~Dick.
\newblock Dynamic stability of electrostatically actuated initially curved
  shallow micro beams.
\newblock {\em Continuum Mech. Thermodyn.}, 22(6):445--468, 2010.

\bibitem{flores2003}
G.~Flores, G.~A. Mercado, and J.~A. Pelesko.
\newblock Dynamics and touchdown in electrostatic {MEMS}.
\newblock In {\em Proc. ASME Design Engineering Technical Conf. and Computers
  and Information in Engineering Conf. and 19th Biennial Conf. on Mechanical
  Vibration and Noise (Chicago, IL, 2--6 September 2003)}, volume~5, pages
  1807--1814. American Society of Mechanical Engineers, 2003.

\bibitem{nielson2006}
G.~N. Nielson and G.~Barbastathis.
\newblock Dynamic pull-in of parallel-plate and torsional electrostatic {MEMS}
  actuators.
\newblock {\em J. Microelectromech. Syst.}, 15(4):811--821, 2006.

\bibitem{leus2008}
V.~Leus and D.~Elata.
\newblock On the dynamic response of electrostatic {MEMS} switches.
\newblock {\em J. Microelectromech. Syst.}, 17(1):236--243, 2008.

\bibitem{joglekar2011}
M.~M. Joglekar and D.~N. Pawaskar.
\newblock Estimation of oscillation period/switching time for electrostatically
  actuated microbeam type switches.
\newblock {\em Int. J. Mech. Sci.}, 53(2):116--125, 2011.

\bibitem{loh2012}
O.~Y. Loh and H.~D. Espinosa.
\newblock Nanoelectromechanical contact switches.
\newblock {\em Nat. Nanotechnol.}, 7(5):283--295, 2012.

\bibitem{Gomez2017}
M.~Gomez, D.~E. Moulton, and D.~Vella.
\newblock Critical slowing down in purely elastic `snap-through' instabilities.
\newblock {\em Nature Phys.}, 13:142--145, 2017.

\bibitem{chaikin}
P.~M. Chaikin and T.~C. Lubensky.
\newblock {\em Principles of condensed matter physics}.
\newblock Cambridge University Press, 1995.

\bibitem{strogatz1989}
S.~H. Strogatz and R.~M. Westervelt.
\newblock Predicted power laws for delayed switching of charge-density waves.
\newblock {\em Phys. Rev. B}, 40(15):10501, 1989.

\bibitem{strogatz2015}
S.~H. Strogatz.
\newblock {\em Nonlinear Dynamics and Chaos}.
\newblock Westview Press, 2014.

\bibitem{castaner1999}
L.~M. Castaner and S.~D. Senturia.
\newblock Speed-energy optimization of electrostatic actuators based on
  pull-in.
\newblock {\em J. Microelectomech. Syst.}, 8(3):290--298, 1999.

\bibitem{veijola1995}
T.~Veijola, H.~Kuisma, J.~Lahdenper{\"a}, and T.~Ryh{\"a}nen.
\newblock Equivalent-circuit model of the squeezed gas film in a silicon
  accelerometer.
\newblock {\em Sens. Actuators A Phys}, 48(3):239--248, 1995.

\bibitem{gupta1996}
R.~K. Gupta, E.~S. Hung, Y.-J. Yang, G.~K. Ananthasuresh, and S.~D. Senturia.
\newblock Pull-in dynamics of electrostatically-actuated beams.
\newblock In {\em Proc. Solid State Sensor and Actuator Workshop}, pages 1--2,
  1996.

\bibitem{rocha2004b}
L.~A. Rocha, E.~Cretu, and R.~F. Wolffenbuttel.
\newblock Pull-in dynamics: analysis and modeling of the transitional regime.
\newblock In {\em Proc. MEMS'04 (Maastricht, The Netherlands, 25--29 January
  2004)}, pages 249--252. IEEE, 2004.

\bibitem{krylov2004}
S.~Krylov and R.~Maimon.
\newblock Pull-in dynamics of an elastic beam actuated by continuously
  distributed electrostatic force.
\newblock {\em J. Vib. Acoust.}, 126(3):332--342, 2004.

\bibitem{nijhuis1999}
M.~H.~H. Nijhuis, T.~G.~H. Basten, Y.~H. Wijnant, H.~Tijdeman, and H.~A.~C.
  Tilmans.
\newblock Transient non-linear response of `pull-in {MEMS} devices' including
  squeeze film effects.
\newblock {\em Proc. Eurosensors XIII (The Hague, The Netherlands, 1999)},
  pages 729--732, 1999.

\end{thebibliography}
\end{document}